\documentclass[journal]{IEEEtran}
\usepackage{cite}

\ifCLASSINFOpdf
  \usepackage[pdftex]{graphicx}
\else
\fi
\usepackage{amsmath}
\usepackage{amssymb}
\usepackage{amsthm}
\usepackage{bm}
\usepackage{algorithm}
\usepackage{algpseudocode}
\ifCLASSOPTIONcompsoc
 \usepackage[caption=false,font=normalsize,labelfont=sf,textfont=sf]{subfig}
\else
 \usepackage[caption=false,font=footnotesize]{subfig}
\fi
\usepackage{url}

\usepackage{physics}
\usepackage{booktabs}
\usepackage{threeparttable}
\usepackage{color}
\usepackage{multirow}

\theoremstyle{definition}

\newtheorem{remark}{Remark}

\newtheorem*{criterion*}{Criterion}

\newcommand{\figref}[1]{Fig. \ref{#1}}
\newcommand{\tabref}[1]{Table \ref{#1}}
\newcommand{\alref}[1]{Algorithm \ref{#1}}
\newcommand{\appref}[1]{Appendix \ref{#1}}
\newcommand{\secref}[1]{Section \ref{#1}}

\begin{document}
%
\title{Beam-Delay Domain Channel Estimation for mmWave XL-MIMO Systems}
%
%
%

\author{Hongwei~Hou,~\IEEEmembership{Graduate~Student~Member,~IEEE}, 
  Xuan~He,~\IEEEmembership{Graduate~Student~Member,~IEEE},\\
  Tianhao~Fang,~\IEEEmembership{Graduate~Student~Member,~IEEE},
  ~Xinping Yi,~\IEEEmembership{Member,~IEEE},\\
  Wenjin~Wang,~\IEEEmembership{Member,~IEEE},
  ~Shi~Jin,~\IEEEmembership{Fellow,~IEEE}
\thanks{}
\thanks{}
\thanks{Manuscript received xxx; revised xxx.}
\thanks{Hongwei Hou, Xuan He, Tianhao Fang, Xinping Yi, Wenjin Wang, and Shi Jin are with the National Mobile Communications Research Laboratory, Southeast University, Nanjing 210096, China (e-mail: hongweihou@seu.edu.cn; hexuan0502@seu.edu.cn; tienhfang@seu.edu.cn; xyi@seu.edu.cn; wangwj@seu.edu.cn; jinshi@seu.edu.cn).}}

%
%

\markboth{}%
{}
%



\maketitle

\begin{abstract}%
  This paper investigates the uplink channel estimation of the millimeter-wave (mmWave) extremely large-scale multiple-input-multiple-output (XL-MIMO) communication system in the beam-delay domain, taking into account the near-field and beam-squint effects due to the transmission bandwidth and array aperture growth. 
  Specifically, we model the sparsity in the delay domain to explore inter-subcarrier correlations and propose the beam-delay domain sparse representation of spatial-frequency domain channels. 
  The independent and non-identically distributed Bernoulli-Gaussian models with unknown prior hyperparameters are employed to capture the sparsity in the beam-delay domain, posing a challenge for channel estimation.
  Under the constrained Bethe free energy minimization framework, we design different structures on the beliefs to develop hybrid message passing (HMP) algorithms, thus achieving efficient joint estimation of beam-delay domain channel and prior hyperparameters. 
  To further improve the model accuracy, the multidimensional grid point perturbation (MDGPP)-based representation is presented, which assigns individual perturbation parameters to each multidimensional discrete grid. 
  By treating the MDGPP parameters as unknown hyperparameters, we propose the two-stage HMP algorithm for MDGPP-based channel estimation, where the output of the initial estimation stage is pruned for the refinement stage for the computational complexity reduction. 
  Numerical simulations demonstrate the significant superiority of the proposed algorithms over benchmarks with both near-field and beam-squint effects.
\end{abstract}%


%
\IEEEpeerreviewmaketitle

\bstctlcite{IEEEexample:BSTcontrol}

\section{Introduction}

\IEEEPARstart{D}{riven} by the demands of higher data rates, both millimeter-wave (mmWave) and extremely large-scale multiple-input multiple-output (XL-MIMO) technologies are expected to play pivotal roles for the future cellular communication systems due to the vast spectrum and high spectral efficiency \cite{6736750, 7959169, han2020channel, 10104065, 10098681}. Accurate channel estimation is crucial for enabling the performance gains of mmWave and XL-MIMO, which is significantly challenged by the near-field and beam-squint effects \cite{7942128, 7414041, wang2018spatial-tsp, mojahedian2022spatial, luo2023beam}.

Due to the extremely large aperture array (ELAA) employed in the mmWave XL-MIMO systems, the Rayleigh distance, which is proportional to the square of the array aperture, becomes comparable to the cell coverage radius \cite{7942128}. 
This fact allows mobile terminal (MTs) and scatterers to appear in the near-field region, thereby leading to the special wave propagation \cite{7414041}.
On the other hand, large bandwidths are typically employed in the mmWave systems, such as the maximum transmission bandwidth of up to $2$ GHz in the fifth-generation (5G) New Radio (NR) specifications \cite{3gpp:38101-2}, which results in a rather low sampling interval.
Therefore, when mmWave is integrated with ELAA, the propagation delay difference over the antenna array will be higher than the sampling interval, resulting in the beam-squint effect \cite{wang2018spatial-tsp, mojahedian2022spatial, luo2023beam}.
With the above analysis, it is shown that the near-field and beam-squint effects in mmWave XL-MIMO systems render its channel model significantly different from that of the conventional massive MIMO (mMIMO) system, which calls for new approaches for accurate channel estimation.

\subsection{Previous Works}
At millimeter-wave frequencies, the channels are typically contributed by very few paths due to the sparse scattering nature \cite{6834753, 8207426, 8424015}, which facilitates compressed sensing techniques to enhance the channel estimation performance.
For conventional mMIMO systems, the orthogonal matching pursuit (OMP)-based channel estimation algorithm is proposed in \cite{7458188}, which exploits beam domain sparsity based on the angle sampling and reconstructs the channel by extracting the dominant paths.
For orthogonal frequency division multiplexing (OFDM) systems, \cite{8323164} and \cite{gonzalez2020wideband} present the simultaneously weighted OMP (SW-OMP) algorithm and subcarrier selection-simultaneous iterative gridless weighted-orthogonal least squares (SS-SIGW-OLS) to exploit the joint sparsity of beam domain channels under different subcarriers.
Furthermore, channel estimation based on beam-delay domain sparsity is proposed in \cite{7961152} to explore the inter-antenna and inter-subcarrier correlations.

Since the near-field effect in XL-MIMO systems destroys the plane wave assumption, the beam domain sparsity based on the angle sampling no longer holds, necessitating the novel beam domain representation. 
Following the spherical wave in the near-field channels, the beam domain sparsity based on the angle-distance sampling, denoted as polar domain, is proposed in \cite{cui2022channel, 10078317}, followed by the OMP algorithm for channel estimation. 
To alleviate the estimation error introduced by discrete sampling, \cite{cui2022channel} presents the polar domain simultaneously iterative gridless weighted (P-SIGW) algorithm, which updates the angle-distance sampling grids based on the gradient descent.
As the scatterers and MTs may be distributed in both far-field and near-field regions, the hybrid-field model is employed in \cite{9940281, 9598863, 10233609, 10233516} for such scenarios, which exploit both angle and angle-distance sampling.

With the transmission bandwidth and array aperture growth, the beam-squint effect becomes more significant, which destroys the joint sparsity of beam domain across subcarriers.
To address this issue, \cite{9399122} and \cite{9444239} propose the pattern of beam domain sparsity variations with subcarriers in far-field channels, followed by the generalized simultaneous OMP (GSOMP) the beam-split pattern detection (BSPD) algorithms.
Furthermore, by exploiting the beam-delay domain sparsity with beam-squint effects, the super resolution-based channel estimation algorithms are proposed in \cite{9351751, 8882325, 8714079}.
Most recently, the channel estimation with both near-field and beam-squint effects is investigated in \cite{cui2023near, elbir2023nba}. 
Specifically, \cite{cui2023near} explores the bilinear patterns of beam domain sparsity variations with subcarriers in the common sensing matrix, facilitating the bilinear patterns detection (BPD)-based channel estimation algorithm.
Besides, the near-field beam-split-aware OMP (NBA-OMP) algorithms is proposed in the \cite{elbir2023nba} based on the GSOMP algorithm, which exploits the frequency selective sensing matrices to explore beam domain sparsity based on angle-distance domain sampling and the joint sparsity across subcarriers. 
The exploration of joint sparsity variations due to beam squint effects in \cite{elbir2023nba, cui2023near} provides one solution for the channel estimation in mmWave XL-MIMO systems, paving the way for subsequent work.

\subsection{Motivations and Contributions}
In mmWave XL-MIMO systems, the near-field and beam-squint effects induce substantial changes in channel models, necessitating more accurate modeling.
Besides, there are significant correlations between subcarriers in the OFDM systems, which are under-explored in such scenarios. 
Finally, the sparse model based on discretized sampling encounters an inherent mismatch with continuous parameters, thereby warranting further investigation into effective mitigation strategies for this issue.

Motivated by the previous works, we investigate the channel estimation for mmWave XL-MIMO systems with near-field and beam-squint effects. The main contributions of this paper are summarized as follows:
\begin{itemize}
  \item By incorporating the near-field and beam-squint effects, we model the channel in the beam-delay domain to explore inter-antenna and inter-subcarrier correlations. In doing so, we end up with the beam-delay domain sparse representation of the spatial-frequency domain channel. 
  To capture the beam-delay domain sparsity, the independent and non-identically distributed Bernoulli-Gaussian models with unknown prior hyperparameters are employed, facilitating the formulation of the channel estimation problem.
  \item To cope with the unknown hyperparameters in the prior models, we introduce the constrained Bethe free energy (BFE) minimization framework, which enables the joint estimation of the beam-delay domain channel and the hyperparameters in the prior model. The hybrid message passing (HMP) algorithm is established by designing different belief structures to estimate beam-delay domain channels and their prior hyperparameters, thereby, spatial-frequency domain channels.
  \item To further improve the model accuracy, we introduce the multidimensional grid point perturbation (MDGPP)-based representation, which assigns individual perturbation parameters to each multidimensional discrete grid and thus mitigates mismatches between continuous parameters and discrete grids. 
  Under the constrained BFE minimization framework, the MDGPP parameters are treated as unknown hyperparameters, facilitating the HMP algorithm for MDGPP-based channel estimation. 
  To reduce the computational complexity, we propose the two-stage HMP algorithm, where the output of the initial estimation stage is pruned for the refinement stage.
\end{itemize}

\textit{Organization}:
The rest of this paper is organized as below. 
\secref{sec:system_model} describes the system model and formulates the channel estimation problem.
In \secref{sec:BFE_CE}, the HMP algorithm is developed based on constrained BFE minimization framework.
The two-stage HMP algorithm for the MDGPP-based channel estimation is proposed in \secref{sec:LCOG_CE}, whose computational complexity is also analyzed.
Numerical simulations in \secref{sec:simulation_result} shows the superiority of the proposed algorithms over the benchmarks. 
Finally, \secref{sec:conclusion} concludes the paper.

\textit{Notation}: 
Throughout this paper, imaginary unit is represented by $j = \sqrt{-1}$. 
Lowercase, bold lowercase, and bold uppercase letters denote scalars, column vectors, and matrices, respectively. 
The transpose, conjugate, and conjugate-transpose operations are represented by the superscripts $(\cdot)^{T}$, $(\cdot)^{*}$, and $(\cdot)^{H}$, respectively.
The Kronecker and Hardmard products are represented by the operator $\otimes$ and $\circ$, respectively.
$[\cdot]_{i}$ and $[\cdot]_{i,j}$ are the $i$-th element of a vector and the $(i, j)$-th element of a matrix, respectively.
$\norm{\cdot}_{2}$ and $\norm{\cdot}_{\infty}$ represent the $\ell_{2}$ and $\ell_{\infty}$ norms, respectively.
$\max\left\{\cdot\right\}$ and $\min\left\{\cdot\right\}$ denote the maximum and minimum operators, respectively.
$\mathsf{E}\left\{\cdot\right\}$ and $\mathsf{V}\left\{\cdot\right\}$ denote the expectation and variance operators, respectively.
$\lfloor \cdot \rfloor$, $\lceil \cdot \rceil$, and $\langle \cdot \rangle$ denote the floor, ceil, and modulo operations, respectively.
The element-wise inverse is represented by the superscripts $(\cdot)^{{\circ}-1}$.
The symbols $\mathbb{C}$ denote the complex number fields.
$\mathcal{CN}(x; {\mu}, {\sigma}^{2})$ denotes the random variable $x$ obeying circularly symmetric complex Gaussian distribution with mean $\mu$ and variance ${\sigma}^{2}$.

\section{System Model and Problem Formulation}\label{sec:system_model}
In this paper, we consider the XL-MIMO-OFDM system in mmWave frequencies, which consists of a BS equipped with a uniform linear array (ULA) of $N$ ($N{\;\gg\;}1$) antennas and $U$ single-antenna MTs.
In the time dulplex division (TDD) systems, the channel is estimated from the uplink pilot symbols. It is assumed that the $N_\text{p}$ pilot subcarriers in each training symbol are equally spaced without overlapping for each MT.
Let $N_\text{all}$ and ${\Delta}f$ denote the number of subcarriers and subcarrier spacing, respectively. Then the transmission bandwidth and pilot subcarrier spacing can be defined as $B = N_\text{all}{\Delta}f$ and ${\Delta}\bar{f} = B / N_\text{p}$, respectively. 

\subsection{Received Signal Model and Channel Model}
In the mmWave XL-MIMO systems, the BS usually adopts hybrid precoding architecture, which makes only a far smaller number of baseband received signals at each training symbol observed.
With the hybrid precoding architecture, we assume that the BS is equipped with $N_{\text{RF}}$ RF chains ($N_{\text{RF}}{\;\ll\;}N$), where the RF chains and the antennas are connected by phase shifters network\cite{7370753}.
Besides, quasi-static channel assumptions are considered in this paper, where BS, MTs and scatters remain nearly unchanged during channel estimation. 
Thus, the received signal $\mathbf{y}_{b, p}$ under the $p$-th pilot subcarrier at the $b$-th training symbol can be expressed as\footnote{Since non-overlapping pilot subcarriers are allocated for different MTs, we can focus on only the single MT and omit the MT index to simplify the expression.}
\begin{equation}
  \mathbf{y}_{b, p} = \mathbf{F}_{b, p}^{H}\mathbf{h}_{p}x_{b, p} + \mathbf{z}_{b, p},
\end{equation}
where $x_{b, p}$ denotes the pilot under the $p$-th pilot subcarrier at the $b$-th training symbol, which can be set to $1$ without loss of generality, and $\mathbf{z}_{b, p}{\;\sim\;}\mathcal{CN}\left(0, {\sigma}_{z}^{2} \mathbf{I} \right)$ denotes the additive white Gaussian nosie (AWGN) under the $p$-th pilot subcarrier.

In the hybrid precoding architecture, the number of measurements required for reliable channel estimation cannot be satisfied with only one pilot symbol, and thus $N_\text{PS}$ pilot symbols are required. Stacking all the received signal under the $p$-th pilot subcarrier in a column vector, we can obtain
\begin{equation}
  \mathbf{y}_{p} = \mathbf{F}_{p}^{H} \mathbf{h}_{p} + \mathbf{z}_{p},
\end{equation}
where $\mathbf{F}_{p} = \left[\mathbf{F}_{0, p}, {\dots}, \mathbf{F}_{N_\text{RX}-1, p}\right]{\;\in\;}\mathbb{C}^{N{\times}M}$ denotes the hybrid precoder and $M = N_\text{RF}N_\text{PS}$ denotes the number of measurements.
Then, all received signals under $N_\text{p}$ pilot subcarriers are stacked in a column vector, so that the received signal $\mathbf{y}$ can be expressed as
\begin{equation}\label{eq:received_signal}
  \mathbf{y} = \bar{\mathbf{F}}^{H}\mathbf{h} + \mathbf{z},
\end{equation}
where $\bar{\mathbf{F}} = \mathsf{blkdiag}\{\mathbf{F}_{0}, {\dots}, \mathbf{F}_{N_\text{p}-1}\}$ denotes the hybrid precoder of all pilot subcarriers, and $\mathbf{h}$ denotes the spatial-frequency domain channel.

We assume that the channel between the target MT and the BS has $L$ paths. Then the channel between the MT and the $n$-th antenna of BS under the $p$-th pilot subcarrier can be expressed as
\begin{equation}\label{eq:scalar_sf_channel}
  h_{p, n} = \sum_{l=0}^{L-1} \frac{{\alpha}_{l}}{r_{l}^{\left(n\right)}} \exp[-j\frac{2{\pi}f_{p}}{c}\left( r_{l}^{\left(n\right)} + c\bar{\tau}_{l} \right)],
\end{equation}
where ${\alpha}_{l}$ denotes the complex gain of $l$-th path, $r_{l}^{\left(n\right)}$ denotes the distance between the $n$-th antenna of BS and the last-hop scatter (LHS)\footnote{For LOS path and NLOS path, the LHS is defined as the MT and the last scatterer before arriving at the antenna array, respectively.} of $l$-th path, $\bar{\tau}_{l}$ denotes the propagation delay between the MT and LHS of $l$-th path, $f_{p} = f_\text{c} + \left(p - N_\text{p}/2\right){\Delta}\bar{f}$ denotes the operating frequency of the $p$-th pilot subcarrier, and $c$ denotes the speed of light.

\begin{figure}[!t]
  \centering
  \includegraphics[width = \linewidth]{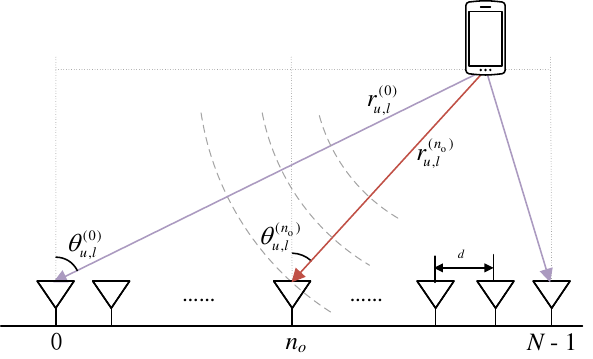}
  \caption{Near-field channel model based on spherical wave.}
  \label{fig:Spherical_Wave}
\end{figure}

As shown in \figref{fig:Spherical_Wave}, the distance between the $n$-th antenna of BS and LHS of the $l$-th path for MT satisfies \cite{7414041}:
\begin{align}\label{eq:distance_approximation}
  r_{l}^{\left(n\right)} &= \sqrt{\left[r_{l}^{\left(n\right)} \cos {\theta}_{l}^{\left(n\right)}\right]^{2} + \left[r_{l}^{\left(n\right)} \sin {\theta}_{l}^{\left(n\right)}\right]^{2}} \nonumber \\
  &\mathop{=}^{\left(a\right)} \sqrt{\left[r_{l}^{\left(n_\text{o}\right)} \cos {\theta}_{l}^{\left(n_\text{o}\right)}\right]^{2} + \left[r_{l}^{\left(n_\text{o}\right)} \sin {\theta}_{l}^{\left(n_\text{o}\right)} + {\Delta}nd\right]^{2}},
\end{align}
where ${\theta}_{l}^{\left(n\right)}$ denotes the physical angle between the $n$-th antenna of ULA and LHS of the $l$-th path, $n_{o}$ denotes the index of reference antenna, ${\Delta}n{\;\triangleq\;}n - n_\text{o}$ denotes the antenna index difference between the $n$-th antenna and the reference antenna, and $\left(a\right)$ follows from the fact that the geometric constraint $r_{l}^{\left(n\right)}\cos {\theta}_{l}^{\left(n\right)}$ is constant, $\forall n, l$. With the Taylor series expansion of $\sqrt{1+x}$ at $x=0$, $r_{l}^{\left(n\right)}$ can be approximated as
\begin{equation}
  r_{l}^{\left(n\right)} \mathop{{\;\approx\;}}^{\left(a\right)} r_{l}^{\left(n_\text{o}\right)} + {\Delta}nd\psi_{l}^{\left(n_\text{o}\right)} + ({\Delta}n)^{2}d^{2}\frac{1-\left[{\psi}_{l}^{\left(n_\text{o}\right)}\right]^{2}}{r_{l}^{n_\text{o}}},
\end{equation}
where ${\psi}_{l}^{\left(n_\text{o}\right)}{\;\triangleq\;}\sin \psi_{l}^{\left(n_\text{o}\right)}$ denotes the angle between the $n$-th antenna of BS and LHS of the $l$-th path, $d = c / (2f_\text{c})$ denotes the half-wavelength antenna spacing, and $\left(a\right)$ is obtained by ignoring the higher order terms of $\Delta{n}d$.

For convenience, $r_{l}{\;\triangleq\;}r_{l}^{\left(n_\text{o}\right)}$ and ${\psi}_{l}{\;\triangleq\;}{\psi}_{l}^{\left(n_\text{o}\right)}$ are regarded as the location parameters of the $l$-th path for MT without causing confusion. Substituting \eqref{eq:distance_approximation} into \eqref{eq:scalar_sf_channel} and stacking the spatial domain channels into vector form, we can obtain
\begin{equation}\label{eq:spatial_channel_p_sc}
  \mathbf{h}_{p} = \sum_{l=0}^{L-1} \bar{\alpha}_{l} e^{-j2{\pi}f_{p}{\tau}_{l}} \mathbf{a}\left({r}_{l}, {\psi}_{l}, f_{p}\right) {\;\circ\;} \mathbf{b}\left({r}_{l}, {\psi}_{l}\right),
\end{equation}
where ${\tau}_{l} = \bar{\tau}_{l} + r_{l} / c$ denotes the propagation delay between the MT and reference antenna at BS, $\mathbf{a}\left(r, {\psi}, f_{p}\right)$ denotes phase differences under the $p$-th pilot subcarrier, which can be defined by
\begin{equation}
  \left[ \mathbf{a}\left({r}, {\psi}, f_{p}\right) \right]_{n}\!=\!\exp\left[ -j\frac{2{\pi}f_{p}}{c}\left( {\Delta}nd{\psi}\!+\!({\Delta}n)^{2}d^{2}\frac{1-{\psi}^{2}}{r} \right)  \right],
\end{equation}
$\mathbf{b}\left(r, {\psi}\right)$ denotes power differences, which can be defined by
\begin{equation}
  \left[ \mathbf{b}\left({r}, {\psi}\right) \right]_{n} = \frac{\varrho}{r + {\Delta}nd\psi+ ({\Delta}n)^{2}d^{2}\frac{1-{\psi}^{2}}{r}},
\end{equation}
where $\varrho$ denotes the normalized factor, such that
\begin{equation}
  \frac{1}{{\varrho}^{2}} = \sum_{n=0}^{N-1} \frac{1}{ \left(r + {\Delta}nd\psi+ ({\Delta}n)^{2}d^{2}\frac{1-{\psi}^{2}}{r}\right)^{2} },
\end{equation}
and the common phase and gain between all antennas independent of ${\Delta}n$ is merged into the path gain $\bar{\alpha}_{l}$. 
For convenience, we define $\mathbf{c}\left(r, {\psi}, f_{p}\right){\;\triangleq\;}\mathbf{a}\left({r}, {\psi}, f_{p}\right) {\;\circ\;} \mathbf{b}\left({r}, {\psi}\right)$ as beam domain steering vector.

In \eqref{eq:spatial_channel_p_sc}, the differences between the mmWave XL-MIMO and mMIMO channel model are two-fold:
\begin{enumerate}
  \item Near-field effect: In mmWave XL-MIMO systems, the MTs and LHSs located within the Rayleigh distance result in the spherical wave propagation \cite{7942128, 7414041}. Moreover, the propagation distance difference causes fluctuations in power attenuation across antennas.
  \item Beam-squint effect: In mmWave XL-MIMO systems, the propagation delay differences on the antenna array are not negligible due to the large bandwidth of mmWave systems, which leads to the frequency-dependent beam domain steering vector \cite{wang2018spatial-tsp}.
\end{enumerate}

To illustrate the deviation from the plane wave assumption due to near-field effects and facilitate subsequent beam domain representations, we can introduce the slope parameter $\eta{\;\triangleq\;}\left(1-{\psi}^{2}\right)/r$ to characterize the rate of instantaneous angle change with ${\Delta}n$, which is shown by
\begin{equation}
  \left[ \mathbf{a}\left({r}, {\psi}, f_{p}\right) \right]_{n} \!=\! \exp\Bigg\{\!-\!j\frac{2{\pi}f_{p}}{c}\Bigg[ {\Delta}nd \Big( \underbrace{{\psi}\!+\!{\Delta}nd\frac{1-{\psi}^{2}}{r}}_{\text{Instantaneous angle}} \Big) \Bigg] \Bigg\}.
\end{equation}
Then, the angle-distance sampling can be converted to the angle-slope sampling by replace the $\left(1-{\psi}^{2}\right)/r$ with ${\eta}$, resulting in the beam domain steering vector $\mathbf{c}\left({\eta}, {\psi}, f_{p}\right)$.

By stacking the spatial channel of all pilot subcarriers, the spatial-frequency channel is expressed as
\begin{equation}\label{eq:spatial_frequency_channel}
  \mathbf{h} = \sum_{l=0}^{L-1} {\alpha}_{l} \left[\mathbf{d}\left({\tau}_{l}\right){\;\otimes\;}\mathbf{1}_{N}\right]{\;\circ\;}\bar{\mathbf{c}}\left({\eta}_{l}, {\psi}_{l}\right),
\end{equation}
where $\mathbf{d}\left({\tau}\right)$ denotes the delay domain steering vector defined as $\left[\mathbf{d}\left({\tau}\right)\right]_{p} = \exp(-j2{\pi}f_{p}{\tau})$, and $\bar{\mathbf{c}}\left({\eta}, {\psi}\right)$ denotes aggregate beam domain steering vector given by
\begin{equation}
  \bar{\mathbf{c}}\left({\eta}, {\psi}\right) = \left[{\mathbf{c}}^{T}\left({\eta}, {\psi}, f_{0}\right), {\dots}, {\mathbf{c}}^{T}\left({\eta}, {\psi}, f_{N_\text{p}-1}\right)\right]^{T}.
\end{equation}

\begin{remark}
  When specific effects and structure are ignored, the conventional mMIMO channel model can be considered as special cases of the proposed channel model.
  \begin{enumerate}
    \item With both near-field and beam-squint effects, but ignoring the correlation across subcarriers, we have $\mathbf{d}(\tau) = \bm{\rho}$, whose elements are uncorrelated. Thus, the channel model in \eqref{eq:spatial_frequency_channel} degenerates to the form in \cite{cui2023near, elbir2023nba}.
    \item With the beam-squint effect only, we have ${\eta}{\;\approx\;}0$, thus the channel model in \eqref{eq:spatial_frequency_channel} degenerates to the form in \cite{8882325, 8714079}.
    \item With the near-field effect only, we have $f_{p}{\;\approx\;}f_\text{c}, \forall p$, thus the channel model in \eqref{eq:spatial_frequency_channel} degenerates to the form in \cite{cui2022channel}.
    \item Without both near-field and beam-squint effects, we have ${\eta}{\;\approx\;}0$ and $f_{p}{\;\approx\;}f_\text{c}, \forall p$, thus channel model in \eqref{eq:spatial_frequency_channel} degenerates to the form in \cite{8323164, 7961152}.
  \end{enumerate}
\end{remark}

\subsection{Beam-Delay Domain Sparse Representation}

From \eqref{eq:spatial_frequency_channel}, the spatial-frequency domain channel is superposed by several paths, resulting in the inter-antenna and inter-subcarrier correlations.
To exploit such correlations, we develop the beam-delay domain sparse representation.
Toward this end, we assume that the angles, slopes and delays for each MT satisfy ${\psi}{\;\in\;}\left[-1, 1\right)$, ${\eta}{\;\in\;}\left[0, {\eta}_{\max}\right]$, and ${\tau}{\;\in\;}\left[0, {\tau}_{\max}\right]$, respectively. The maximum slope parameter ${\eta}_{\max} = 1/r_{\min}$, where $r_{\min}$ denotes the minimum service distance, and ${\tau}_{\max}$ denotes the maximum allowable delay, which is determined by the cyclic prefix length and the array aperture under the beam-squint effect.
Then, the angle, slope, and delay domain are sampled uniformly at intervals ${\psi}_{\Delta} = 2/N$, ${\eta}_{\Delta}$\footnote{The sampling interval ${\eta}_{\Delta}$ is determined by the discrete Fresnel function with the coherence threshold \cite{cui2022channel}.}, and ${\tau}_{\Delta}=1/(N_\text{p}{\Delta}\bar{f})$ to obtain the sampling sets, which can be defined by \eqref{eq:sampling_set_angle}, \eqref{eq:sampling_set_slope}, and \eqref{eq:sampling_set_delay}, respectively.
\begin{subequations}
  \begin{equation}\label{eq:sampling_set_angle}
    \Upsilon_{\text{an}} = \left\{ \bar{\psi}_{k_\text{an}}: -1 + \left(k_\text{an} + \frac{1}{2}\right){\psi}_{\Delta}, k_\text{an}{\;\in\;}\mathcal{K}_\text{an} \right\}, 
  \end{equation}
  \begin{equation}\label{eq:sampling_set_slope}
    \Upsilon_{\text{sl}} = \left\{ \bar{\eta}_{k_\text{sl}}: {\eta}_{\min} + \left(k_\text{sl} + \frac{1}{2}\right){\eta}_{\Delta}, k_\text{sl}{\;\in\;}\mathcal{K}_\text{sl} \right\}, 
  \end{equation}
  \begin{equation}\label{eq:sampling_set_delay}
    \Upsilon_{\text{de}} = \left\{ \bar{\tau}_{k_\text{de}}: \left(k_\text{de} + \frac{1}{2}\right){\tau}_{\Delta}, k_\text{de}{\;\in\;}\mathcal{K}_\text{de} \right\},
  \end{equation}
\end{subequations}
where $\mathcal{K}_\text{x} {\;\triangleq\;} \left\{ 0, 1, {\dots}, K_\text{x} - 1\right\}, \text{x}{\;\in\;}\left\{\text{an}, \text{sl}, \text{de}\right\}$ denotes the sampling point index set, $K_\text{an} = \lceil 2/{\psi}_{\Delta} \rceil$, $K_\text{an} = \lceil {\eta}_{\max}/{\eta}_{\Delta} \rceil$, and $K_\text{de} = \lceil {\tau}_{\max}/{\tau}_{\Delta} \rceil$ denote the number of sampling points in angle, slope, and delay domains, respectively. 
Since the sampling intervals decrease as $N$ and $N_\text{p}$ increase, the spatial-frequency domain channel is well approximated as a linear combination of basis functions determined by angle-slope-delay tuples $\left( \bar{\psi}_{k_\text{an}}, \bar{\eta}_{k_\text{sl}}, \bar{\tau}_{k_\text{de}} \right)$ for sufficiently large $N$ and $N_\text{p}$.

Therefore, with the pre-defined sampling set of the angles, slopes, and delays, \eqref{eq:spatial_frequency_channel} can be approximated as
\begin{equation}\label{eq:sf_channel_asd_summation}
  \mathbf{h} {\;\approx\;} \sum_{{k}_\text{an}} \sum_{{k}_\text{sl}} \sum_{{k}_\text{de}} {\beta}_{k_\text{an}, k_\text{sl}, k_\text{de}} \mathbf{u}\left( \bar{\psi}_{k_\text{an}}, \bar{\eta}_{k_\text{sl}}, \bar{\tau}_{k_\text{de}} \right),
\end{equation}
where $\mathbf{u}\left( \bar{\psi}_{k_\text{an}}, \bar{\eta}_{k_\text{sl}}, \bar{\tau}_{k_\text{de}} \right)$ denotes the angle-slope-delay domain basis function defined as
\begin{equation}
  \mathbf{u}\left( \bar{\psi}_{k_\text{an}}, \bar{\eta}_{k_\text{sl}}, \bar{\tau}_{k_\text{de}} \right) = \left[\mathbf{d}\left(\bar{\tau}_{k_\text{de}}\right){\;\otimes\;}\mathbf{1}_{N}\right]{\;\circ\;}\bar{\mathbf{c}}\left(\bar{\eta}_{k_\text{sl}}, \bar{\psi}_{k_\text{an}}\right),
\end{equation}
and ${\beta}_{k_\text{an}, k_\text{sl}, k_\text{de}}$ denotes the complex gain of the path with respect to angle-slope-delay tuple $\left( \bar{\psi}_{k_\text{an}}, \bar{\eta}_{k_\text{sl}}, \bar{\tau}_{k_\text{de}} \right)$, which is given by
\begin{equation}
  {\beta}_{k_\text{an}, k_\text{sl}, k_\text{de}} = \begin{cases}
    {\alpha}_{l}, &\text{if}\;\left( \bar{\psi}_{k_\text{an}}, \bar{\eta}_{k_\text{sl}}, \bar{\tau}_{k_\text{de}} \right) = \left({\psi}_{l}, {\eta}_{l}, {\tau}_{l}\right) \\
    0, &\text{otherwise}
  \end{cases}.
\end{equation}

Reformulating \eqref{eq:sf_channel_asd_summation} into matrix form and assuming that the spatial-frequency domain channels can be represented perfectly, we can obtain
\begin{equation}\label{eq:asd_channel_representation}
  \mathbf{h} = \mathbf{U}\bm{\beta},
\end{equation}
where $\mathbf{U}{\;\in\;}\mathbb{C}^{NN_\text{p}{\times}K_\text{an}K_\text{sl}K_\text{de}}$ and $\bm{\beta}{\;\in\;}\mathbb{C}^{K_\text{an}K_\text{sl}K_\text{de}{\times}1}$ denote the beam-delay domain transformation matrix and the beam-delay domain channel defined as
\begin{equation}
  \left[\mathbf{U}\right]_{:, k_\text{de}K_\text{an}K_\text{sl} + k_\text{an}K_\text{sl} + k_\text{sl}} = \mathbf{u}\left( \bar{\psi}_{k_\text{an}}, \bar{\eta}_{k_\text{sl}}, \bar{\tau}_{k_\text{de}} \right),
\end{equation}
and
\begin{equation}
  \left[\bm{\beta}\right]_{k_\text{de}K_\text{an}K_\text{sl} + k_\text{an}K_\text{sl} + k_\text{sl}} = {\beta}_{k_\text{an}, k_\text{sl}, k_\text{de}},
\end{equation}
respectively. Without causing confusion, $(k_\text{an}, k_\text{sl}, k_\text{de})$ maps to the simple index $k{\;\in\;}\{0,1,{\dots},K-1\}$ in a one-to-one manner, which satisfies 
\begin{equation}\label{eq:index_mapping}
  k_\text{de} = \lfloor \frac{k}{K_\text{an}K_\text{sl}} \rfloor, k_\text{an} = \lfloor \frac{\langle k \rangle_{K_\text{an}K_\text{sl}}}{K_\text{sl}} \rfloor, k_\text{an} = \langle k \rangle_{K_\text{sl}}, 
\end{equation}
and $K = K_\text{an}K_\text{sl}K_\text{de}$.

\subsection{Channel Estimation Problem Formulation}
With the proposed sparse representation, the channel estimation problem is to estimate the beam-delay domain channel and ultimately the spatial-frequency domain channel based on the received signal. The MMSE channel estimator can be expressed as the \textit{a posteriori} mean \cite{kay1993fundamentals}, given by
\begin{equation}\label{eq:MMSE_estimator_asd}
  \bm{\beta}^{\star} = \int \bm{\beta} \mathsf{Pr}\left(\bm{\beta}, \mathbf{s} \mid \mathbf{y}\right) \mathrm{d} \bm{\beta} \mathrm{d} \mathbf{s},
\end{equation}
where $\mathbf{s}{\;\triangleq\;}\mathbf{G}\bm{\beta}$ and $\mathbf{G}{\;\triangleq\;}\bar{\mathbf{F}}^{H}\mathbf{U}$ denote the auxiliary vector and measurement matrix, respectively, $\mathsf{Pr}\left(\bm{\beta}, \mathbf{s} \mid \mathbf{y}\right)$ denotes the posterior probability density function (PDF) of $\bm{\beta}$ and $\mathbf{s}$. 

From the received signal model and the beam-delay domain channel representation, $p\left(\bm{\beta}, \mathbf{s}\mid \mathbf{y}\right)$ can be expressed as
\begin{equation}
  \mathsf{Pr}\left(\bm{\beta},  \mathbf{s} \mid \mathbf{y}\right) {\;\propto\;} \mathsf{Pr}\left(\mathbf{y} \mid \mathbf{s} \right) \mathsf{Pr}\left(\mathbf{s} \mid \bm{\beta} \right) \mathsf{Pr}\left(\bm{\beta}\right),
\end{equation}
where $\mathsf{Pr}\left(\mathbf{y} \mid \mathbf{s} \right)$, $\mathsf{Pr}\left(\mathbf{s} \mid \bm{\beta} \right)$, and $\mathsf{Pr}\left(\bm{\beta}\right)$ denote received noise model, baseband transfer model, and beam-delay domain channel prior model, respectively.

Since AWGN channels are employed, the conditional PDF $\mathsf{Pr}\left( \mathbf{y} \mid \mathbf{s} \right)$ can be detailed as 
\begin{equation}
  \mathsf{Pr}\left( \mathbf{y} \mid \mathbf{s} \right) = \prod_{m}\prod_{p} \underbrace{\mathcal{CN}\left(y_{m, p}; s_{m, p}, {\sigma}_{z}^{2} \right)}_{{\;\triangleq\;}\mathsf{Pr}\left(y_{m, p} \mid s_{m, p}\right)},
\end{equation}
where $y_{m, p}$ and $s_{m, p}$ denote the $\left(mN_\text{p}+p\right)$-th element of $\mathbf{y}$ and $\mathbf{s}$, respectively. Then the conditional PDF $\mathsf{Pr}\left( \mathbf{s} \mid \bm{\beta} \right)$ can be factorized as
\begin{equation}
  \mathsf{Pr}\left( \mathbf{s} \mid \bm{\beta} \right) =  \prod_{m}\prod_{p} \underbrace{\delta\left(s_{m, p} - \mathbf{g}_{m, p}^{(\text{r})} \bm{\beta} \right)}_{{\;\triangleq\;}\mathsf{Pr}\left(s_{m, p} \mid \bm{\beta}\right)},
\end{equation}
where $\mathbf{g}_{m, p}^{(\text{r})}$ denotes the $\left(mN_\text{p}+p\right)$-th row of $\mathbf{G}$.

With the assumption of wide-sense stationary uncorrelated scattering Rayleigh fading channel, the elements of $\bm{\beta}$ are statistically independent for sufficiently large $N$ and $N_\text{p}$ \cite{yang2022channel}.
Moreover, the sparse nature of mmWave channels is considered, implying that only a few number of elements in $\bm{\beta}$ have significant power, with most elements being zero.
Thus, we can model $\bm{\beta}$ as an independent and non-identically distributed Bernoulli-Gaussian distribution, and the prior PDF $\mathsf{Pr}\left(\bm{\beta}; \bm{\omega}\right)$ controlled by hyperparameters $\bm{\omega}$ can be decomposed into
\begin{equation}
  \mathsf{Pr}\left(\bm{\beta}; \bm{\omega}\right) = \prod_{k} \underbrace{\left[ \left(1-{\lambda}\right){\delta}\left({\beta}_{k}\right) + {\lambda} \mathcal{CN}\left({\beta}_{k}; 0, {\chi}_{k}\right) \right]}_{{\;\triangleq\;}\mathsf{Pr}\left({\beta}_{k}; {\lambda}, {\chi}_{k}\right)},
\end{equation}
where $\bm{\omega} = \left[{\lambda}, {\chi}_{0}, {\dots}, {\chi}_{K-1}\right]$, ${\lambda}$ denotes the sparsity level, and ${\chi}_{k}$ denotes the power of ${\beta}_{k}$ when ${\beta}_{k}$ is nonzero. 

Building on the probabilistic modeling, the MMSE estimation of spatial-frequency domain channel can be given by 
\begin{equation}
  \mathbf{h}^{\star} = \mathbf{U}\bm{\beta}^{\star},
\end{equation}
due to the commutability of the MMSE estimator over linear transformations \cite{kay1993fundamentals}.

\section{Constrained BFE Minimization-based Channel Estimation}\label{sec:BFE_CE}
The MMSE estimator in \eqref{eq:MMSE_estimator_asd} requires the multi-dimensional integrals of $\mathsf{Pr}\left(\bm{\beta}, \mathbf{s} \mid \mathbf{y}; \bm{\omega}\right)$, which introduces an unacceptable complexity in large-scale systems. 
In this section, we approximate $\mathsf{Pr}\left(\bm{\beta}, \mathbf{s} \mid \mathbf{y}; \bm{\omega}\right)$ with structured trial beliefs based on the Bethe method, and then develop the HMP algorithm to achieve efficient joint estimation of beam-delay domain channel and prior hyperparameters.

\subsection{Constrained BFE Minimization Framework}
Following the VFE minimization framework \cite{jordan1999introduction, winn2005variational, bishop2006pattern}, we approximate $\mathsf{Pr}\left(\bm{\beta}, \mathbf{s} \mid \mathbf{y}; \bm{\omega}\right)$ by the trial belief $\mathsf{b}\left(\bm{\beta}, \mathbf{s}; \bm{\omega}\right)$, which can be obtained by minimizing the Kullback-Leibler (KL) divergence of $b\left(\bm{\beta}, \mathbf{s}; \bm{\omega}\right)$ and $\mathsf{Pr}\left(\bm{\beta}, \mathbf{s} \mid \mathbf{y}; \bm{\omega}\right)$. The KL divergence minimization problem can be expressed by
\begin{align}
  \hat{\mathsf{b}}\left(\bm{\beta}, \mathbf{s}; \bm{\omega}\right) &= \arg\min_{b{\in}\mathcal{Q}} D\left[ \mathsf{b}\left(\bm{\beta}, \mathbf{s}; \bm{\omega}\right) \| \mathsf{Pr}\left(\bm{\beta}, \mathbf{s} \mid \mathbf{y}; \bm{\omega}\right) \right] \nonumber \\
  &= \arg\min_{b{\in}\mathcal{Q}} \mathcal{F}_{V} + \ln Z\left(\mathbf{y}\right),
\end{align}
where $\mathcal{Q}$ denotes the constrained set of $\mathsf{b}\left(\bm{\beta}, \mathbf{s}; \bm{\omega}\right)$, $D\left[\cdot \| \cdot \right]$ denotes the KL divergence, $\mathcal{F}_{V}$ denotes the VFE defined by
\begin{equation}
  \mathcal{F}_{V} {\;\triangleq\;} \iint \mathsf{b}\left(\bm{\beta}, \mathbf{s}; \bm{\omega}\right) \ln \frac{\mathsf{b}\left(\bm{\beta}, \mathbf{s}; \bm{\omega}\right)}{\mathsf{Pr}\left(\bm{\beta}, \mathbf{s}, \mathbf{y}; \bm{\omega}\right)} \mathrm{d}\bm{\beta}\mathrm{d}\mathbf{s},
\end{equation}
where $\mathsf{Pr}\left(\bm{\beta}, \mathbf{y}; \bm{\omega}\right)$ denotes the joint PDF, and $-\ln Z\left(\mathbf{y}\right){\;\triangleq\;}-\ln p\left(\mathbf{y}\right)$ denotes the Helmholtz free energy independent of $\bm{\beta}$. Thus, the KL divergence minimization is equivalent to the VFE minimization.

The VFE minimization is still intractable in large-scale systems, which motivates the constrained set design of $\mathsf{b}\left(\bm{\beta}; \bm{\omega}\right)$. To balance the exactness and tractability, the Bethe method is adopted to design the constrained set and convert the VFE minimization into the BFE minimization.
Following Bayes rule and the probabilistic model, the joint PDF can be factorized as
\begin{align}
  \mathsf{Pr}&\left(\bm{\beta}, \mathbf{s}, \mathbf{y}; \bm{\omega}\right) \nonumber \\
  &=\!\prod_{m}\prod_{p} \underbrace{ \mathsf{Pr}\left(y_{m, p}\!\mid\!s_{m, p}\right)}_{{\;\triangleq\;}f_{y, m, p}} \underbrace{\mathsf{Pr}\left( s_{m, p}\!\mid\!\bm{\beta}\right)}_{{\;\triangleq\;}f_{s, m, p}} \prod_{k} \underbrace{ \mathsf{Pr}\left({\beta}_{k}; {\lambda}, {\chi}_{k}\right)}_{{\;\triangleq\;}f_{\beta, k}},
\end{align}
where $f_{y, m, p}$, $f_{s, m, p}$ and $f_{\beta, k}$ denote the factor nodes. Based on the Bethe method \cite{1459044, 9351786, liu2021sparse}, we introduce auxiliary beliefs $\mathsf{b}_{s, m, p}\left(s_{m, p}\right)$, $\mathsf{b}_{s \bm{\beta}, m, p}\left(s_{m, p}, \bm{\beta}\right)$ and $\mathsf{b}_{{\beta} {\lambda} {\chi}, k}\left({\beta}_{k}, {\lambda}, {\chi}_{k}\right)$ for factor nodes $f_{y, m, p}$, $f_{s, m, p}$ and $f_{{\beta}, k}$, respectively. Similarly, we introduce auxiliary beliefs $\mathsf{q}_{\lambda}\left({\lambda}\right)$, $\mathsf{q}_{{\chi}, k}\left({\chi}_{k}\right)$, $\mathsf{q}_{{\beta}, k}\left({\beta}_{k}\right)$ and $\mathsf{q}_{s, m, p}\left(s_{m, p}\right)$ for variable nodes ${\lambda}$, ${\chi}_{k}$, ${\beta}_{k}$ and $s_{m, p}$, respectively. Following this, the trial belief and the BFE can be expressed as
\begin{equation}
  \mathsf{b}\left(\bm{\beta}, \mathbf{s}; \bm{\omega}\right) = \frac{\prod\limits_{m}\prod\limits_{p} \mathsf{b}_{s, m, p} \mathsf{b}_{s \bm{\beta}, m, p}\prod\limits_{k} \mathsf{b}_{{\beta} {\lambda} {\chi}, k}}{\prod\limits_{m}\prod\limits_{p} \mathsf{q}_{s, m, p}\mathsf{q}_{\lambda}^{K}\prod\limits_{k} \mathsf{q}_{\chi, k}\mathsf{q}_{\beta, k}^{MN_\text{p}}}, 
\end{equation}
and
\begin{align}\label{eq:BFE}
  \mathcal{F}_{B} =& \sum_{m} \sum_{p} \left\{ D\left[ \mathsf{b}_{s,m, p} \| f_{y, m, p} \right] + D\left[ \mathsf{b}_{s \bm{\beta}, m, p} \| f_{s, m, p} \right] \right\}\nonumber \\
  & + \sum_{k} D\left[ \mathsf{b}_{{\beta} {\lambda} {\chi}, k} \| f_{\beta, k}\right] + \sum_{m} \sum_{p} H[\mathsf{q}_{s, m, p}]   \nonumber \\
  & + \sum_{k} MN_\text{p}H\left[\mathsf{q}_{\beta, k}\right] + KH[\mathsf{q}_{\lambda}] + H[\mathsf{q}_{\chi, k}],
\end{align}
respectively, where $H\left[\cdot\right]$ denotes the entropy.

Inspired by the variational message-passing algorithm \cite{winn2005variational}, we can factorize complicated beliefs into the product of simple beliefs by exploiting the slow-varying characteristic of hyperparameters, which can be expressed as
\begin{equation}\label{eq:asd_channel_hyparameter_belief}
  \mathsf{b}_{{\beta} {\lambda} {\chi}, k}\left({\beta}_{k}, {\lambda}, {\chi}_{k}\right) = \mathsf{b}_{\beta, k}\left({\beta}_{k}\right) \mathsf{b}_{\lambda}\left({\lambda}\right) \mathsf{b}_{\chi, k}\left({\chi}_{k}\right), 
\end{equation}
where $\mathsf{b}_{\beta, k}\left({\beta}_{k}\right)$, $\mathsf{b}_{\lambda}\left({\lambda}\right)$ and $\mathsf{b}_{\chi, k}\left({\chi}_{k}\right)$ denote the auxiliary belief with respect to ${\beta}_{k}$, ${\lambda}$ and ${\chi}_{k}$, respectively.
Furthermore, the hyperparameter beliefs can be constrained  by the Dirac delta function, which is given as
\begin{subequations}\label{eq:hyparameter_belief}
  \begin{equation}
    \mathsf{b}_{\lambda}\left({\lambda}\right) = \delta\left({\lambda} - {\lambda}_\text{true}\right),
  \end{equation}
  \begin{equation}
    \mathsf{b}_{\chi, k}\left({\chi}_{k}\right) = \delta\left({\chi}_{k} - {\chi}_{k, \text{true}}\right), 
  \end{equation}
\end{subequations}
where ${\lambda}_\text{true}{\;\in\;}\left(0, 1\right)$ and ${\chi}_{k, \text{true}}{\;\geq\;}0$ denote the ground-truth hyperparameters.

To guarantee the global dependence of factor nodes, the auxiliary beliefs of factor and variable nodes in Bethe method should satisfy marginal consistency constraints (MCCs), which is intractable for continuous random variables. To address this issue, the MCCs can be relaxed to mean and variance consistency constraints (MVCCs), which can be given by
\begin{subequations}\label{eq:MVCC_NR}
  \begin{equation}
    \mathsf{E}\left\{\bm{\beta} \mid \mathsf{q}_{\beta}\right\} = \mathsf{E}\left\{\bm{\beta} \mid \mathsf{b}_{\beta}\right\} = \mathsf{E}\left\{\bm{\beta} \mid \mathsf{b}_{s \bm{\beta}, m, p}\right\}, \forall m, p, 
  \end{equation}
  \begin{equation}
    \mathsf{V}\left\{\bm{\beta} \mid \mathsf{q}_{\beta}\right\} = \mathsf{V}\left\{\bm{\beta} \mid \mathsf{b}_{\beta}\right\} = \mathsf{V}\left\{\bm{\beta} \mid \mathsf{b}_{s \bm{\beta}, m, p}\right\}, \forall m, p,
  \end{equation}
  \begin{equation}
    \mathsf{E}\left\{\mathbf{s} \mid \mathsf{q}_{s}\right\} = \mathsf{E}\left\{\mathbf{s} \mid \mathsf{b}_{s}\right\} = \mathsf{E}\left\{ \mathbf{s} \mid \mathsf{b}_{s \bm{\beta}}\right\},
  \end{equation}
  \begin{equation}
    \mathsf{V}\left\{\mathbf{s} \mid \mathsf{q}_{s}\right\} = \mathsf{V}\left\{\mathbf{s} \mid \mathsf{b}_{s}\right\} = \mathsf{V}\left\{ \mathbf{s} \mid \mathsf{b}_{s \bm{\beta}}\right\},
  \end{equation}
\end{subequations}
where the beliefs without index subscripts denote the vector version of the scalar beliefs. The variance consistency constraints with respect to $\bm{\beta}$ and $\mathsf{b}_{s \bm{\beta}}$ are further relaxed as \eqref{eq:RVCC} by averaging, which can be given by
\begin{equation}\label{eq:RVCC}
  \mathsf{V}\left\{\bm{\beta} \mid \mathsf{q}_{\beta}\right\} = \mathsf{V}\left\{\bm{\beta}\mid \mathsf{b}_{\beta}\right\} = \frac{1}{MN_{\text{p}}} \sum_{m} \sum_{p} \mathsf{V}\left\{\bm{\beta} \mid \mathsf{b}_{s \bm{\beta}, m, p}\right\}.
\end{equation}

Finally, the KL divergence minimization problem can be converted into the constrained BFE minimization problem, which is expressed by
\begin{equation}\label{eq:constrained_BFE_minimization}
  \min_{b{\in}\mathcal{Q}} {\ } \mathcal{F}_{B}, {\quad}\text{s.t.} {\ } \eqref{eq:asd_channel_hyparameter_belief}, \eqref{eq:hyparameter_belief}, \eqref{eq:MVCC_NR}, \eqref{eq:RVCC}.
\end{equation}

\subsection{HMP Algorithm for Channel Estimation}

\begin{algorithm}[!t]
  \caption{HMP Algorithm}
  \label{alg:FPI_CE}
  \begin{algorithmic}[1]
    \Require {$\mathbf{y}$ (received signal), ${\sigma}_{z}^{2}$ (noise variance).}
    \Ensure {Spatial-frequency channel $\hat{\mathbf{h}} = \mathbf{U} \hat{\bm{\beta}}$.}
    \State {Initialize $\hat{\bm{\beta}}, \bm{\sigma}_{\beta}^{2}, \bm{\varsigma}^{{\beta}, b_{s\bm{\beta}}}, \bm{\xi}^{s, b_{s}}$}
    \For {$t = 1, {\dots}, T$}
    \State {$\bm{\varsigma}^{s, b_{s}} = ( {| {\mathbf{G}} |^{{\circ}2}} ( \bm{\varsigma}^{\beta, b_{s\bm{\beta}}} )^{{\circ}-1} )^{{\circ}-1} $} \label{algstate:smp_start}
    \State {$\bm{\mu}^{s, b_{s}} = -{\bm{\xi}^{s, b_{s}}}{\;\circ\;}(\bm{\varsigma}^{s, b_{s}})^{{\circ}-1} + \mathbf{G}\hat{\bm{\beta}}$}
    \State {$\begin{aligned}
        \mathsf{b}_{s}{\;\propto\;}&\mathcal{CN}\left(\mathbf{s}; \mathbf{y}, {\sigma}_{z}^{2}\mathbf{I}\right) {\cdot} \nonumber \\
          &\mathcal{CN}\left(\mathbf{s}; \bm{\mu}^{s, b_{s}}, -\mathsf{diag}\{\bm{\varsigma}^{s, b_{s}}\}^{{\circ}-1}\right)
      \end{aligned}$}
    \State {Obtain $\hat{\mathbf{s}}$ and $\bm{\sigma}_{s}$ following $\mathsf{b}_{s}$.}
    \State {$\bm{\varsigma}^{s, b_{s\bm{\beta}}} = -\bm{\sigma}_{s}^{{\circ}-2} - \bm{\varsigma}^{s, b_{s}}$}
    \State {$\bm{\mu}^{s, b_{s\bm{\beta}}} = {\bm{\xi}^{s, b_{s}}}{\;\circ\;}(\bm{\varsigma}^{s, b_{s\bm{\beta}}})^{{\circ}-1} + \hat{\mathbf{s}}$}
    \State {${\Delta}\bm{\mu} = \bm{\mu}^{s, b_{s\bm{\beta}}}-\bm{\mu}^{s, b_{s}}$}
    \State {$\bm{\xi}^{s, b_{s}} = ((\bm{\varsigma}^{s, b_{s}})^{{\circ}-1}+(\bm{\varsigma}^{s, b_{s\bm{\beta}}})^{{\circ}-1})^{{\circ}-1}{\Delta}\bm{\mu}$}
    \State {$\bm{\pi}^{s, b_{s}} = \bm{\varsigma}^{s, b_{s}}{\;\circ\;}(1 + \bm{\varsigma}^{s, b_{s}} {\;\circ\;} \bm{\sigma}_{s}^{2})$} \label{algstate:smp_end}
    \State {$\bm{\pi}^{\beta, b_{\beta}} = |\mathbf{G}^{H}|^{{\circ}2}\bm{\pi}^{s, b_{s}}$}\label{algstate:betak_start_infor}
    \State {$\bm{\varsigma}^{\beta, b_{\beta}}=((\bm{\pi}^{\beta, b_{\beta}})^{{\circ}-1} - (MN_\text{p}\bm{\varsigma}^{\beta, b_{s\bm{\beta}}})^{{\circ}-1})^{{\circ}-1}$}
    \State {$\bm{\mu}^{\beta} = (\bm{\varsigma}^{\beta, b_{\beta}})^{{\circ}-1} {\;\circ\;} (\mathbf{G}^{H}\bm{\xi}^{s, b_{s}}) + \hat{\bm{\beta}}$}
    \State {$\begin{aligned}
      \mathsf{b}_{\beta}{\;\propto\;}&[ (1-\hat{\lambda}){\delta}(\bm{\beta}) + \hat{\lambda} \mathcal{CN}(\bm{\beta}; 0, \mathsf{diag}\{\hat{\bm{\chi}}\}) ] {\cdot}\nonumber \\
      &\mathcal{CN}( \bm{\beta}; \bm{\mu}^{\beta}, -\mathsf{diag}\{\bm{\varsigma}^{\beta, b_{\beta}}\}^{{\circ}-1})
    \end{aligned}$}
    \State {Obtain $\hat{\bm{\beta}}$ and $\bm{\sigma}_{\beta}$ following $\mathsf{b}_{{\beta}}$.}
    \State {$\bm{\varsigma}^{\beta, b_{s\bm{\beta}}} = - \bm{\sigma}_{\beta}^{{\circ}-2} - \bm{\varsigma}^{\beta, b_{\beta}}/(MN_\text{p})$}\label{algstate:betak_end_infor}
    \State {Update $\hat{\chi}_{k}, \forall k$ by \eqref{eq:chi_k_learning} and $\hat{\lambda}$ by \eqref{eq:lambda_learning}.}
    \EndFor
  \end{algorithmic}
\end{algorithm}

The optimization problem in \eqref{eq:constrained_BFE_minimization} can be solved by the Lagrange multiplier method \cite{boyd2004convex}, which sets the derivative of the Lagrangian function with respect to the auxiliary beliefs to zero. 
Since the auxiliary beliefs of the hyperparameters can be completely factorized in \eqref{eq:asd_channel_hyparameter_belief}, their MCCs can be satisfied constantly and thus can be ignored in the Lagrangian function. Substituting constraints \eqref{eq:asd_channel_hyparameter_belief} and \eqref{eq:hyparameter_belief} into $\mathsf{b}_{{\beta}{\lambda}{\chi}, k}$, the Lagrangian function of \eqref{eq:constrained_BFE_minimization} can be expressed as
\begin{equation}
  \mathcal{L}_{\text{B}} = \mathcal{F}_{\text{B, HP}} + \mathcal{L}_{\text{M}} + \mathcal{L}_{\text{V}}, 
\end{equation}
where $\mathcal{F}_{\text{B, HP}}$ denotes the BFE with factorized hyperparameter constraints, which can be obtained by replacing the KL-divergence $D[\mathsf{b}_{{\beta}{\lambda}{\chi}, k} \| f_{\beta, k}]$ with $D[\mathsf{b}_{{\beta}, k} \| \hat{f}_{\beta, k}]$ in \eqref{eq:BFE}, and $\hat{f}_{\beta, k} {\;\triangleq\;}\mathsf{Pr}({\beta}_{k}; {\lambda}_{\text{true}}, {\chi}_{k, \text{true}})$, $\mathcal{L}_{\text{M}}$ and $\mathcal{L}_{\text{V}}$ defined in \eqref{eq:MVCC_LF} denote the subfunctions of mean consistency constraints and variance consistency constraints, respectively, $\bm{\xi}^{s, b_{s}}$, $\bm{\xi}^{s, b_{s\bm{\beta}}}$, $\bm{\xi}^{{\beta}, b_{{\beta}}}$, and $\bm{\xi}_{m, p}^{{\beta}, b_{s\bm{\beta}}}$ denote the Lagrangian multiplier for mean consistency constraints, $\bm{\varsigma}^{s, b_{s}}$, $\bm{\varsigma}^{s, b_{s\bm{\beta}}}$, $\bm{\varsigma}^{{\beta}, b_{{\beta}}}$ and $\bm{\varsigma}^{{\beta}, b_{{\beta}}}$ denote the Lagrangian multiplier for variance consistency constraints.

\begin{figure*}[!t]
  \normalsize
  \begin{subequations}\label{eq:MVCC_LF}
    \begin{align}\label{eq:mean_CC_LF}
      \mathcal{L}_{\text{M}} =& 2\Re\left\{ \left(\bm{\xi}^{s, b_{s}}\right)^{H} \left[ \mathsf{E}\left\{\mathbf{s} \mid \mathsf{q}_{s}\right\} - \mathsf{E}\left\{\mathbf{s} \mid \mathsf{b}_{s}\right\} \right]  \right\} + 2\Re\left\{ \left(\bm{\xi}^{s, b_{s\bm{\beta}}}\right)^{H} \left[ \mathsf{E}\left\{\mathbf{s} \mid \mathsf{q}_{s}\right\} - \mathsf{E}\left\{\mathbf{s} \mid \mathsf{b}_{s \bm{\beta}}\right\} \right] \right\}  \nonumber \\
      &+ 2\Re\left\{ \left(\bm{\xi}^{{\beta}, b_{{\beta}}}\right)^{H} \left[\mathsf{E}\left\{\bm{\beta} \mid \mathsf{q}_{\beta}\right\} - \mathsf{E}\left\{\bm{\beta} \mid \mathsf{b}_{\beta}\right\}\right] \right\} + \sum_{m}\sum_{p} 2\Re\left\{ \left(\bm{\xi}_{m, p}^{{\beta}, b_{s\bm{\beta}}}\right)^{H} \left[\mathsf{E}\left\{\bm{\beta} \mid \mathsf{q}_{\beta}\right\} - \mathsf{E}\left\{\bm{\beta} \mid \mathsf{b}_{s \bm{\beta}, m, p}\right\}\right] \right\},
    \end{align}
    \vspace*{-2pt}
    \begin{align}\label{eq:variance_CC_LF}
      \mathcal{L}_{\text{V}} =& \left(\bm{\varsigma}^{s, b_{s}}\right)^{T} \left[ \mathsf{V}\left\{\mathbf{s} \mid \mathsf{q}_{s}\right\} - \mathsf{V}\left\{\mathbf{s} \mid \mathsf{b}_{s}\right\} \right] + \left(\bm{\varsigma}^{s, b_{s\bm{\beta}}}\right)^{T} \left[ \mathsf{V}\left\{\mathbf{s} \mid \mathsf{q}_{s}\right\} - \mathsf{V}\left\{\mathbf{s} \mid \mathsf{b}_{s \bm{\beta}}\right\} \right] \nonumber \\
      &+ \left(\bm{\varsigma}^{{\beta}, b_{{\beta}}}\right)^{T} \left[\mathsf{V}\left\{\bm{\beta} \mid \mathsf{q}_{\beta}\right\} - \mathsf{V}\left\{\bm{\beta} \mid \mathsf{b}_{\beta}\right\}\right] + \left(\bm{\varsigma}^{{\beta}, b_{{\beta}}}\right)^{T} \left[ MN_\text{p}\mathsf{V}\left\{\bm{\beta} \mid \mathsf{q}_{\beta}\right\} - \sum_{m}\sum_{p}\mathsf{V}\left\{\bm{\beta} \mid \mathsf{b}_{s \bm{\beta}, m, p}\right\}\right].
    \end{align}
 \end{subequations}
  \hrulefill
  \vspace*{4pt}
\end{figure*}

Due to space limitations, the stationary-point equations are provided in \appref{app:FP_CBFE}. Building on the derived equations, the resulting HMP algorithm for channel estimation is summarized in \alref{alg:FPI_CE}, where the adaptive damping based on the BFE evaluation \cite{vila2015adaptive} is introduced to improve the convergence.

\begin{remark}
  In the constrained BFE minimization problem, we leverage the properties of different beliefs to design hybrid constraints, thus incorporating the existing message passing algorithms into \alref{alg:FPI_CE}. It is noteworthy that the Bernoulli-Gaussian distribution is only one of the prior models, and other specific forms of $\mathsf{b}_{\beta}$ can be obtained by different prior models, such as Bernoulli-Gaussian-Mixture \cite{vila2013expectation} and Laplacian \cite{bellili2019generalized} distributions.
\end{remark}

\section{MDGPP-Based Channel Estimation}\label{sec:LCOG_CE}
Due to the finite number of antennas and pilot subcarriers in the practical system, the sampling sets fail to match the continuous parameters in the channel perfectly \cite{cui2022channel, 8882325}, resulting in the modeling inaccuracy \cite{5710590, 6576276}. 
To address this issue, we introduce the MDGPP-based representation and propose the two-stage HMP algorithm based on the constrained BFE minimization framework.

\subsection{MDGPP-Based Channel Representation}
\begin{figure}[!t]
  \centering
	\subfloat[Conventional perturbations]{\includegraphics[width = 0.45\linewidth]{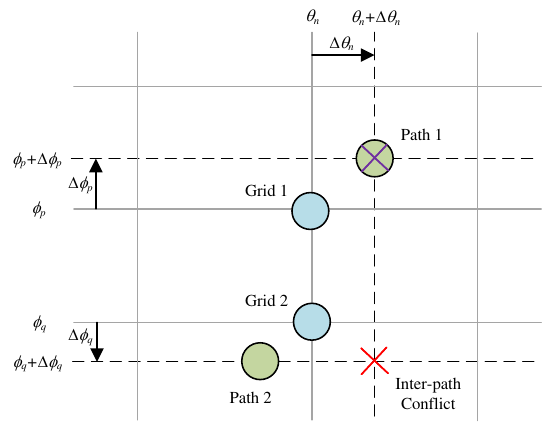}%
		\label{fig:OG_Perturbation_GridLine}}
	\subfloat[MDGPP]{\includegraphics[width = 0.45\linewidth]{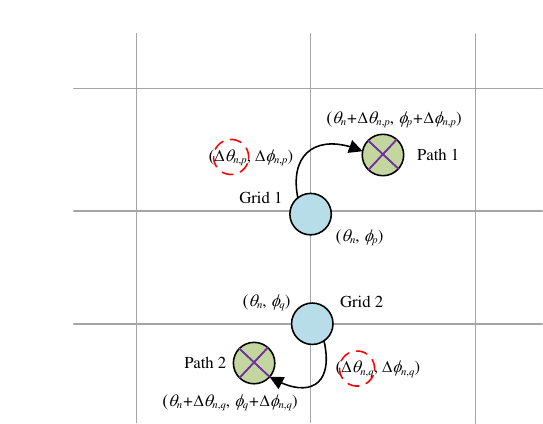}%
		\label{fig:OG_Perturbation_GridPoint}}
	\caption{Conventional perturbations and MDGPP of the two-dimensional example (Green circle: ground-truth parameter, blue circle: grids without perturbation, puple/red cross: matched/mismatched grids with perturbation).}
	\label{fig:OG_Perturbation}
\end{figure}

To eliminate the gap between the sampling sets and the continuous parameters, the spatial-frequency domain channel can be decorrelated by the MDGPP-based channel basis functions, which is expressed as
\begin{equation}
  \mathbf{h} {\;\approx\;} \sum_{k} {\beta}_{k} \tilde{\mathbf{u}}_{k}\left({\Delta}{\psi}_{k}, {\Delta}{\eta}_{k}, {\Delta}{\tau}_{k}\right),
\end{equation}
where ${\Delta}{\psi}_{k}{\;\in\;}\left[-{\psi}_{\Delta}/2, {\psi}_{\Delta}/2\right)$, ${\Delta}{\eta}_{k}{\;\in\;}\left[-{\eta}_{\Delta}/2, {\eta}_{\Delta}/2\right)$, and ${\Delta}{\tau}_{k}{\;\in\;}\left[-{\tau}_{\Delta}/2, {\tau}_{\Delta}/2\right)$ denote the perturbation parameters in angle, slope, and delay of the $k$-th sampling tuple, respectively, ${\beta}_{k}$ denotes the complex gain of the $k$-th sampling tuple, and $\tilde{\mathbf{u}}_{k}\left({\Delta}{\psi}_{k}, {\Delta}{\eta}_{k}, {\Delta}{\tau}_{k}\right)$  denote the perturbed basis function of the $k$-th sampling tuple defined as
\begin{align}
  \tilde{\mathbf{u}}_{k} \left({\Delta}{\psi}_{k}, {\Delta}{\eta}_{k}, {\Delta}{\tau}_{k}\right) = & \left[\mathbf{d}\left(\bar{\tau}_{k_\text{de}} + {\Delta}{\tau}_{k}\right){\;\otimes\;}\mathbf{1}_{N}\right] \nonumber \\
  &{\;\circ\;}\bar{\mathbf{c}}\left(\bar{\eta}_{k_\text{sl}} + {\Delta}{\eta}_{k}, \bar{\psi}_\text{an} + {\Delta}{\psi}_{k}\right),
\end{align}
where the index tuple $\left({k}_\text{an}, {k}_\text{sl}, {k}_\text{de}\right)$ and the simple index $k$ satisfies the mapping rule \eqref{eq:index_mapping}.
The proposed MDGPP-based representation assigns individual perturbation parameters to each beam-delay domain sampling grid, while the conventional perturbation assigns common perturbation parameters based on grid line constraints \cite{liu2020angular}. 
To visualize advantages of the MDGPP-based representation, the schematic of the two-dimensional example is shown in \figref{fig:OG_Perturbation}, where the perturbations are shown on the $(\theta, \phi)$ parameter.
It can be observed that, with conventional perturbations, the perturbation of ``Grid 1'' for matching ``Path 1'' causes the mismatch between ``Grid 2'' and ``Path 2'' due to the common perturbation parameter ${\Delta}{\theta}_{n}$, which can be avoided by assigning different perturbation parameters ${\Delta}{\theta}_{n, p}$ and ${\Delta}{\theta}_{n, q}$ in the MDGPP.  

\begin{remark}
  The conventional perturbation can be considered as a special case of the MDGPP. Specifically, when the MDGPP parameters exhibits the specific structure as
  \begin{subequations}
    \begin{equation}
      \Delta\bm{\psi} = \overline{\Delta\bm{\psi}} {\;\otimes\;} \mathbf{1}_{K_\text{sl}} {\;\otimes\;} \mathbf{1}_{K_\text{de}},
    \end{equation}
    \begin{equation}
      \Delta\bm{\eta} = \mathbf{1}_{K_\text{an}} {\;\otimes\;} \overline{\Delta\bm{\eta}} {\;\otimes\;} \mathbf{1}_{K_\text{de}},
    \end{equation}
    \begin{equation}
      \Delta\bm{\tau} = \mathbf{1}_{K_\text{an}} {\;\otimes\;} \mathbf{1}_{K_\text{sl}} {\;\otimes\;} \overline{\Delta\bm{\tau}},
    \end{equation}
  \end{subequations}
  the MDGPP-based representation degenerates to the form in \cite{liu2020angular}, where $\Delta\bm{\psi}$, $\Delta\bm{\eta}$ and $\Delta\bm{\tau}$ denote vectors of MDGPP parameters in the angle, slope and delay domains, respectively, and $\overline{\Delta\bm{\psi}}$, $\overline{\Delta\bm{\eta}}$ and $\overline{\Delta\bm{\tau}}$ denote vectors of conventional perturbation parameters in the angle, slope and delay domains, respectively.
\end{remark}

Benefiting from MDGPP-based representations, each path can be associated with the sampling tuple in $\Upsilon$ which is closest to the path parameters, and the deviations are described by the MDGPP parameters.
By stacking the perturbed basis function into matrix form, the spatial-frequency domain channel can be given by the bilinear form as
\begin{equation}
  \mathbf{h} = \tilde{\mathbf{U}}\left( \Delta\bm{\psi}, \Delta\bm{\eta}, \Delta\bm{\tau} \right)\bm{\beta}, 
\end{equation}
where $\tilde{\mathbf{U}}\left( \Delta\bm{\psi}, \Delta\bm{\eta}, \Delta\bm{\tau} \right)$ denotes the perturbed transformation maxtrix with perturbation parameter vector $\Delta\bm{\psi}$, $\Delta\bm{\eta}$, and $\Delta\bm{\tau}$. 
To balance the exactness and complexity, the perturbed transformation maxtrix is approximated by the Taylor series \cite{apostol1974mathematical}, which can be expressed as
\begin{align}
  \tilde{\mathbf{U}}\left( \Delta\bm{\psi}, \Delta\bm{\eta}, \Delta\bm{\tau} \right) = & \mathbf{U} + \mathbf{U}_{\psi}\mathsf{diag}\left\{ \Delta\bm{\psi} \right\} +  \nonumber \\
  & \mathbf{U}_{\eta}\mathsf{diag}\left\{ \Delta\bm{\eta} \right\} + \mathbf{U}_{\tau}\mathsf{diag}\left\{ \Delta\bm{\tau} \right\},
\end{align}
where $\mathbf{U}_{\psi}$, $\mathbf{U}_{\eta}$, and $\mathbf{U}_{\tau}$ denote the first-order partial derivatives of $\mathbf{U}$ with respect to angle, slope and delay.

\subsection{Two-Stage HMP Algorithm for MDGPP-based Channel Estimation}
\begin{figure*}[!t]
  \centering
  \includegraphics[width = 0.8\linewidth]{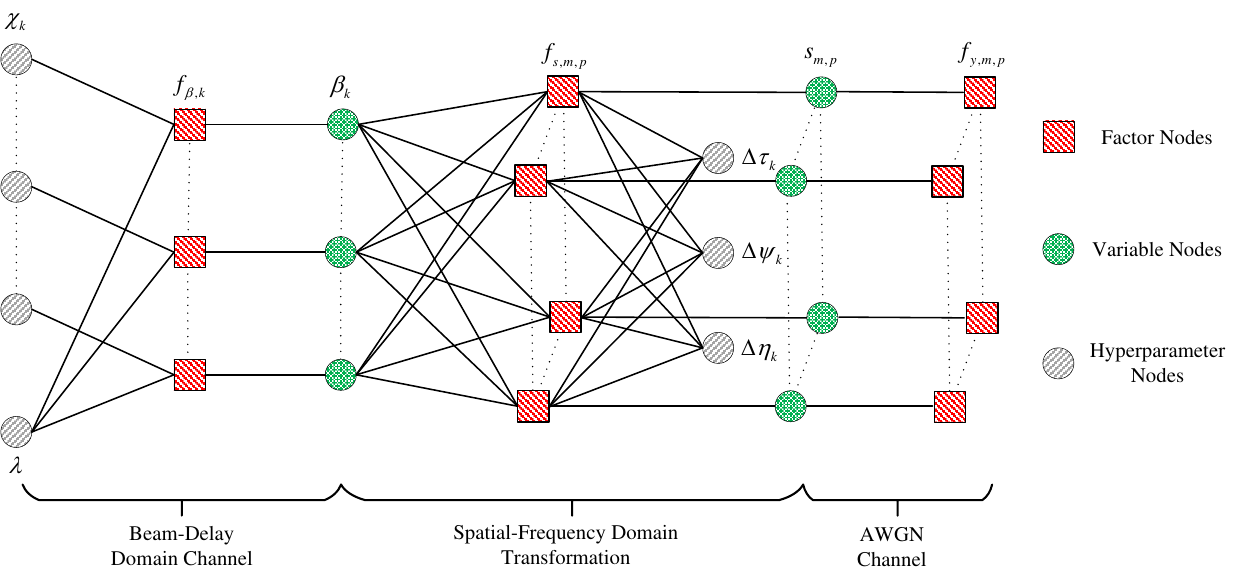}
  \caption{Factor graph representation of the MDGPP-based channel estimation.}
  \label{fig:Factor_Graph_OG}
\end{figure*}

Following the proposed representation, the channel estimation problem aims to estimate both beam-delay domain channel and the MDGPP paramters, leading to the structured bilinear model.
To tackle this problem, the joint PDF with the MDGPP paramters is factorized as
\begin{align}
  \mathsf{Pr}&\left(\mathbf{y}, \mathbf{s}, \bm{\beta}, \Delta\bm{\psi}, \Delta\bm{\eta}, \Delta\bm{\tau}\right) = p\left( \mathbf{y} \mid \mathbf{s} \right) \mathsf{Pr}\left(\bm{\beta}\right)  \nonumber \\
  & \mathsf{Pr}\left( \mathbf{s} \mid \bm{\beta}, \Delta\bm{\psi}, \Delta\bm{\eta}, \Delta\bm{\tau}\right) \mathsf{Pr}\left(\Delta\bm{\psi}\right)\mathsf{Pr}\left(\Delta\bm{\eta}\right)\mathsf{Pr}\left(\Delta\bm{\tau}\right), 
\end{align}
where the MDGPP parameters are assumed to be mutually independent of each other and beam-delay domain channels. 
With the independent assumptions, the prior PDF of MDGPP parameters can be fully factorized to $\mathsf{Pr}\left(\Delta{\psi}_{k}\right)$, $\mathsf{Pr}\left(\Delta{\eta}_{k}\right)$, and $\mathsf{Pr}\left(\Delta{\tau}_{k}\right)$, which can be denoted by $f_{\Delta{\psi}, k}$, $f_{\Delta{\eta}, k}$, and $f_{\Delta{\tau}, k}$, respectively. The conditional PDF can be factorized as
\begin{align}
  \mathsf{Pr} &\left( \mathbf{s} \mid \bm{\beta}, \Delta\bm{\psi}, \Delta\bm{\eta}, \Delta\bm{\tau}\right)\nonumber \\
  &=\prod_{m}\prod_{p} \underbrace{\mathsf{Pr}\left(s_{m, p} \mid \bm{\beta}, \Delta\bm{\psi}, \Delta\bm{\eta}, \Delta\bm{\tau}\right)}_{\tilde{f}_{s, m, p}}.
\end{align}
Building on factorizations of the joint PDF, the factor graph of MDGPP-based channel estimation is shown in \figref{fig:Factor_Graph_OG}. 
We introduce auxiliary beliefs $\mathsf{b}_{\Delta{\psi}, k}\left({\Delta}{\psi}_{k}\right)$, $\mathsf{b}_{\Delta{\eta}, k}\left({\Delta}{\eta}_{k}\right)$, $\mathsf{b}_{\Delta{\tau}, k}\left({\Delta}{\tau}_{k}\right)$, and $\mathsf{b}_{s\bm{\beta}\Delta\bm{\psi}\Delta\bm{\eta}\Delta\bm{\tau}, m, p}\left(s_{m, p}, \bm{\beta}, \Delta\bm{\psi}, \Delta\bm{\eta}, \Delta\bm{\tau}\right)$ for factor nodes $f_{\Delta{\psi}, k}$, $f_{\Delta{\eta}, k}$, $f_{\Delta{\tau}, k}$, and $\tilde{f}_{s, m, p}$, respectively. Similarly, the auxiliary beliefs  $\mathsf{q}_{\Delta{\psi}, k}\left({\Delta}{\psi}_{k}\right)$, $\mathsf{q}_{\Delta{\eta}, k}\left({\Delta}{\eta}_{k}\right)$, and $\mathsf{q}_{\Delta{\tau}, k}\left({\Delta}{\tau}_{k}\right)$ are introduced for variable nodes $\Delta{\psi}_{k}$, $\Delta{\eta}_{k}$, and $\Delta{\tau}_{k}$, respectively.
To guarantee the tractability of the structured bilinear inference problem, the complicated beliefs $\mathsf{b}_{s\bm{\beta}\Delta\bm{\psi}\Delta\bm{\eta}\Delta\bm{\tau}, m, p}$ can be factorized as
\begin{equation}
  \mathsf{b}_{s\bm{\beta}\Delta\bm{\psi}\Delta\bm{\eta}\Delta\bm{\tau}, m, p} = \mathsf{b}_{s\bm{\beta}, m, p} \prod_{k}\left(\mathsf{b}_{\Delta{\psi}, k}\mathsf{b}_{\Delta{\eta}, k}\mathsf{b}_{\Delta{\tau}, k}\right),
\end{equation}
where $\mathsf{b}_{\Delta{\psi}, k}$, $\mathsf{b}_{\Delta{\eta}, k}$, and $\mathsf{b}_{\Delta{\tau}, k}$ are further constrained by the Dirac delta function with ground-truth perturbation parameters ${\Delta}{\psi}_{k, \text{true}}{\in}\left[-{\psi}_{\Delta}/2, {\psi}_{\Delta}/2\right)$, ${\Delta}{\eta}_{k, \text{true}}{\in}\left[-{\eta}_{\Delta}/2, {\eta}_{\Delta}/2\right)$, and ${\Delta}{\tau}_{k, \text{true}}{\in}\left[-{\tau}_{\Delta}/2, {\tau}_{\Delta}/2\right)$.

With the formulation above, the perturbation parameters can be acquired efficiently by the constrained BFE minimization framework.
Compared with \alref{alg:FPI_CE}, the HMP algorithm for MDGPP-based channel estimation differ in the update of perturbation parameters, which is derived in \appref{app:FP_CBFE_OG}. 
The update rule of angle domain perturbation parameters is
\begin{equation}\label{eq:obj_problem_psi}
  \Delta\bm{\psi}^{\star} = \arg\min_{\Delta\bm{\psi}} \Delta\bm{\psi}^{T} \mathbf{P}_{\psi} \Delta\bm{\psi} - 2 \mathbf{u}_{\psi}^{T}\Delta\bm{\psi},
\end{equation}
where $\mathbf{P}_{\psi}$ and $\mathbf{u}_{\psi}$ can be defined by
\begin{equation}
  \mathbf{P}_{\psi} {\;\triangleq\;} \left(\mathbf{U}_{\psi}^{H} \bar{\mathbf{F}} \bar{\mathbf{F}}^{H} \mathbf{U}_{\psi}\right)^{\ast} {\;\circ\;} \left( \hat{\bm{\beta}}\hat{\bm{\beta}}^{H} + \bm{\Sigma}_{\beta} \right), 
\end{equation}
and
\begin{align}
  \mathbf{u}_{\psi} {\;\triangleq\;} &\Re\left\{ \mathsf{diag}\left\{ \hat{\bm{\beta}}^{\ast} \right\} \mathbf{U}_{\psi}^{H} \bar{\mathbf{F}} \left( \mathbf{y} - \bar{\mathbf{F}}^{H} \mathbf{U}_{\backslash {\psi}}\hat{\bm{\beta}} \right)\right\} \nonumber \\
  &- \Re\left\{\mathsf{diag}\left\{ \mathbf{U}_{\psi}^{H} \bar{\mathbf{F}} \bar{\mathbf{F}}^{H} \mathbf{U}_{\backslash {\psi}} \bm{\Sigma}_{\beta} \right\} \right\},
\end{align}
where $\mathbf{U}_{\backslash {\psi}} {\;\triangleq\;}  \mathbf{U} + \mathbf{U}_{\eta}\mathsf{diag}\left\{ \Delta\bm{\eta} \right\} + \mathbf{U}_{\tau}\mathsf{diag}\left\{ \Delta\bm{\tau} \right\}$, and $\bm{\Sigma}_{\beta}$ denotes the posterior covariance matrix whose diagonal elements are $\bm{\sigma}_{\beta}^{{\circ}2}$ in \alref{alg:FPI_CE}.
To solve the above optimization problem, the closed-form solution without bounding constraints is obtained and then checked whether the bounding constraints are satisfied.
When the closed-form solution does not satisfy the bounding constraints, the perturbation parameters are adjusted sequentially by
\begin{equation}\label{eq:seq_update_psi}
  \Delta{\psi}_{k}^{\star}\!=\!\mathrm{Limit}\left\{ \frac{\left[\mathbf{u}_{\psi}\right]_{n}\!-\!\left[\mathbf{P}_{\psi}\right]_{\backslash n, n}^{T} \left[\Delta\bm{\psi}\right]_{\backslash n} }{ \left[\mathbf{P}_{\psi}\right]_{n, n}},\!-\!\frac{{\psi}_{\Delta}}{2},\!\frac{{\psi}_{\Delta}}{2} \right\},
\end{equation}
where $\mathrm{Limit}\left\{x, a, b\right\}{\;\triangleq\;}\min\left\{ \max\left\{x, a\right\}, b\right\}$.
Similarly, the slope and delay domain perturbation parameters $\Delta\bm{\eta}$ and $\Delta\bm{\tau}$ can be obtained in the same manner.

However, the closed-form solution of the MDGPP parameters requires matrix inversion, which has a prohibitively high computational complexity in large-scale systems. 
Motivated by the mmWave channel sparsity, the two-stage HMP algorithm for MDGPP-based channel estimation can be proposed in \alref{alg:FPI_LCOGCE}, which involves the initial estimation and refinement stages.
In the first stage, the initial estimation of the beam-delay domain channel is obtained by \alref{alg:FPI_CE} without model perturbations.
Building upon the initial estimation, the elements with energies below a certain threshold are pruned to significantly reduce the number of perturbation parameters to be estimated next.
In the second stage, the pruned beam-delay domain channel and its corresponding perturbation parameters are iteratively acquired.

\begin{algorithm}[!t]
  \caption{Two-Stage HMP Algorithm}
  \label{alg:FPI_LCOGCE}
  \begin{algorithmic}[1]
    \Require {$\mathbf{y}$, ${\sigma}_{z}^{2}$, $E_\text{th}$ (energy threshold).}
    \Ensure {$\hat{\mathbf{h}} = \mathbf{U} \hat{\bm{\beta}}$.}
    \State {Execute \alref{alg:FPI_CE} to obtain the initial estimation of beam-delay domain channel $\hat{\bm{\beta}}_{\text{ini}}$.}
    \State {Determine the unpurned indices based on the energy criterion $\mathcal{S}_\text{ref} = \{ k: [\hat{\bm{\beta}}_{\text{ini}}]_{k} > E_\text{th} \| \hat{\bm{\beta}}_{\text{ini}} \|_{\infty} \} $.}
    \State {Prune the initial estimation of beam-delay domain channel $\hat{\bm{\beta}}_{\text{ref}} = [\hat{\bm{\beta}}_{\text{ini}}]_{\mathcal{S}_\text{ref}}$, and $K_\text{ref}{\;\triangleq\;}\mathsf{card}(\mathcal{S}_\text{ref})$.}
    \State {Initialize $\Delta\bm{\psi}$, $\Delta\bm{\eta}$, and $\Delta\bm{\tau}$ to $\mathbf{0}$.}
    \For {$t = 1, {\dots}, T_\text{ref}$}
    \State {Construct $\tilde{\mathbf{G}} = \bar{\mathbf{F}}^{H}\tilde{\mathbf{U}}$ based on $\Delta\bm{\psi}$, $\Delta\bm{\eta}$, and $\Delta\bm{\tau}$.}
    \State {Obtain pruned measurement matrix $\mathbf{G}_{\text{ref}} = [\tilde{\mathbf{G}}]_{:, \mathcal{S}_\text{ref}}$.}    
    \State {Replace $\hat{\bm{\beta}}$ and $\mathbf{G}$ by $\hat{\bm{\beta}}_{\text{ref}}$ and $\mathbf{G}_{\text{ref}}$, respectively.}
    \State {Execute Line \ref{algstate:smp_start} - Line \ref{algstate:betak_end_infor} in \alref{alg:FPI_CE}.}
    \For {$k = 0, 1, {\dots}, K_\text{ref} - 1$}
    \State {Update $\hat{\chi}_{k}$ by \eqref{eq:chi_k_learning} and $\hat{\lambda}$ by \eqref{eq:lambda_learning}.}
    \State {Update $\Delta{\psi}_{k}$, $\Delta{\eta}_{k}$, and $\Delta{\tau}_{k}$.}
    \EndFor
    \EndFor
  \end{algorithmic}
\end{algorithm}

\subsection{Computational Complexity}
The per-iteration computational complexity of \textbf{\alref{alg:FPI_CE}} is $\mathcal{O}(MN_\text{p}K_\text{an}K_\text{sl}K_\text{de})$, which are dominated by matrix-vector multiplications. 
For \textbf{\alref{alg:FPI_LCOGCE}}, the per-iteration computational complexities in the initial estimation stage and the refinement stage are $\mathcal{O}(MN_\text{p}K_\text{an}K_\text{sl}K_\text{de})$ and $\mathcal{O}(MN_\text{p}K_\text{ref}^{2} + K_\text{ref}^{3})$, where the latter is dominated by the perturbation parameter acquisitions.
Since the size of beam-delay domain channel after pruning can be assumed to keep the same order of magnitude as path number $L$ ($L{\;\ll\;}MN_\text{p}$), we have $K_\text{ref}{\;\ll\;}MN_\text{p}$, such that the per-iteration computational complexity in the refinement stage can be approximated as $\mathcal{O}(MN_\text{p}K_\text{ref}^{2})$.
Therefore, the total computational complexity of \textbf{\alref{alg:FPI_LCOGCE}} is about $\mathcal{O}( MN_\text{p}K_\text{an}K_\text{sl}K_\text{de}T_\text{ini} + MN_\text{p}K_\text{ref}^{2}T_\text{ref})$, where $T_\text{ini}$ and $T_\text{ref}$ denote the number of iterations for the initial estimation stage and refinement stage, respectively.
It can also be found that $K_\text{ref}$ controls the tradeoff between the computational complexity and performance. Specifically, a larger $K_\text{ref}$ leads to higher complexity, while smaller $K_\text{ref}$ results in performance degradations since the true beam-delay domain channel components will be pruned.

It can be observed that the exploration of inter-subcarrier correlations increases the computational complexity but delivers significant performance gains, as verified in subsequent simulations. 
Since the proposed algorithms are inherently compatible with the prior models, the further reduction of the computational complexity is feasible by exploiting the prior information, which can be acquired by the sophisticated statistical channel state information estimation \cite{lu20232d} and channel knowledge map \cite{zeng2021toward}. This reduction unfolds two aspects:
\begin{itemize}
  \item The reduction in unknown variables: With the prior information, the support acquisition of beam-delay domain channels is avoided, which enables the reduction of the unknown variables to the order of the number of paths.
  \item The reduction in observations: The received signals are highly correlated with support-aided beam-delay domain channels. To eliminate such redundancy, the row-orthogonalization \cite{fan2017generalized} can be introduced to reduce the effective observations to the order of unknown variables based on an equivalent BFE representation.
\end{itemize}
Furthermore, the subsequent simulations also indicate that the proposed algorithms requires fewer pilot symbols than benchmarks, enabling a further reduction in computational complexity while maintaining performance.

\section{Numeical Simulations}\label{sec:simulation_result}
Unless otherwise specified, the simulation configurations follow the system model in \secref{sec:system_model} with the paramters summarized in \tabref{tab:simulation_configuration}. 
In the subsequent simulations, the SNR is defined as the received SNR defined as $\text{SNR}\;(\text{dB}) = 10\log_{10}(\mathsf{E}\{ \|\bar{\mathbf{F}}^{H}\mathbf{h}\|_{2}^{2} \} / \mathsf{E}\{ \|\mathbf{z}\|_{2}^{2} \})$.

\begin{table}[!t]
  \caption{Scenario Paramters}
  \label{tab:simulation_configuration}
  \centering
  \begin{tabular}{cc}
  \toprule
  \textbf{Paramter} & \textbf{Value} \\
  \midrule
  Carrier Frequency & $f_{\text{c}} = 30$ GHz \\
  Pilot Subcarrier Spacing & ${\Delta}\bar{f} = 25$ MHz \\
  Number of BS Antennas & $N = 256$ \\
  Number of BS RF Chains & $N_\text{RF} = 16$ \\
  Number of Pilot Subcarriers & $N_\text{p} = 64$ \\
  Number of Pilot Symbols & $ N_\text{PS} = 8 $ \\
  Minimum Service Distance & $r_{\min} = 3$ m \\
  Number of Paths & $L = 6$ \\
  \bottomrule
  \end{tabular}
\end{table}

To demonstrate the superiority of the proposed algorithms, we select the following excellent channel estimation algorithms as benchmarks:
\begin{itemize}
  \item \textbf{P-SIGW} \cite{cui2022channel}: This channel estimation algorithm is designed for near-field channels without beam squint effect, which models the beam domain sparsity with sampling in the angle and distance doamin.
  \item \textbf{GSOMP} \cite{9399122}: This channel estimation algorithm is designed for far-field channels with beam squint effect, which employs frequency-dependent sensing matrices.
  \item \textbf{BPD} \cite{cui2023near}, \textbf{NBA-OMP} \cite{elbir2023nba}: These channel estimation algorithm are designed for near-field channels with beam squint effect, which captures the joint sparsity pattern across subcarriers by bilinear pattern and employs frequency-dependent sensing matrices, respectively.
\end{itemize}
For the comparison with the proposed algorithms and benchmarks, the normalized mean square error (NMSE) of the spatial-frequency channels is adopted as the performance metric, which is defined by:
\begin{equation}
  \text{NMSE}\;(\text{dB}) = 10\log_{10}\left\{ \frac{1}{D} \sum_{d=0}^{D-1} \frac{\|\hat{\mathbf{h}}^{(d)} - \mathbf{h}^{(d)}\|_{2}^{2}}{\| \mathbf{h}^{(d)} \|_{2}^{2}} \right\},
\end{equation}
where $D$ denotes the numebr of simulations, $\mathbf{h}^{(d)}$ and $\hat{\mathbf{h}}^{(d)}$ denote the ground-truth and estimated spatial-frequency channels in the $d$-th simulation, respectively.

\begin{figure}[!t]
  \centering
  \includegraphics[width = \linewidth]{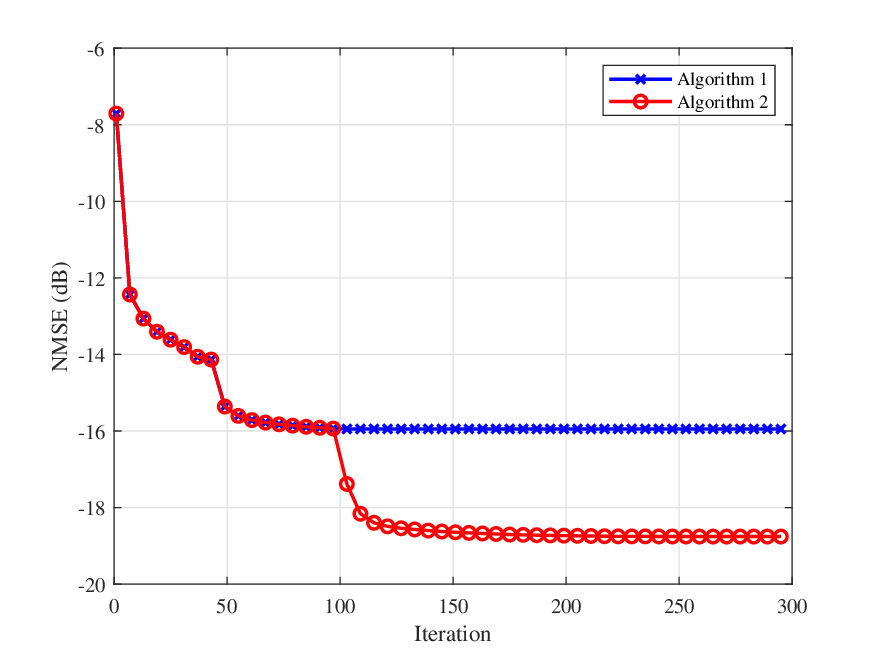}
  \caption{The convergence performance of the proposed algorithms.}
  \label{fig:Convergence}
\end{figure}

\begin{figure}[!t]
  \centering
  \includegraphics[width = \linewidth]{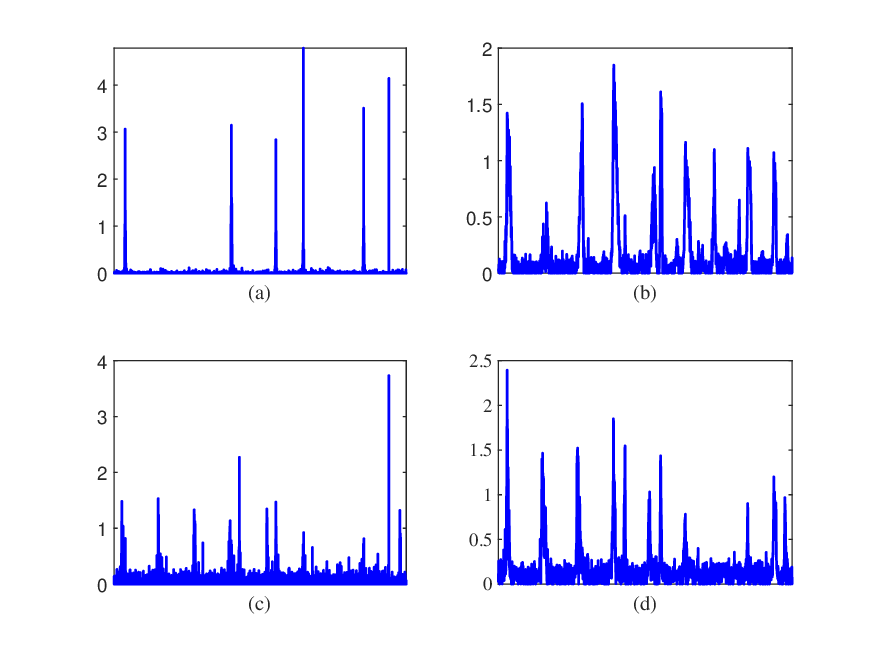}
  \caption{Beam-delay domain channel power of the on-grid case: (a) This work, (b) Without near-field effect, (c) Without beam-squint effect, (d) Without both near-field and beam-squint effects.}
  \label{fig:sparse_representation}
\end{figure}

The convergence performance comparison of \textbf{Algorithm 1} and \textbf{Algorithm 2} at the SNR of $15$dB is provided in \figref{fig:Convergence}. It can be seen that the MDGPP-based representation introduced in \textbf{Algorithm 2} significantly reduces the NMSE from $-15.94$dB to $-18.75$dB.
Since both \textbf{Algorithm 1} and \textbf{Algorithm 2} have rapid decline rate of NMSE in the beginning iterations, we can terminate the algorithm iterations earlier to reduce the computational complexity based on the performance requirements of practical systems. To show the effectiveness of the proposed beam-delay domain sparse representation, the beam-delay domain channel power of the on-grid case estimated from \textbf{Algorithm 1} without specific effects are provided in \figref{fig:sparse_representation}. The proposed beam-delay domain sparse representation can decorrelate the spatial-frequency domain channel effectively, while ignoring the specific effects induces significant energy leakage.

\begin{figure}[!t]
  \centering
  \includegraphics[width = \linewidth]{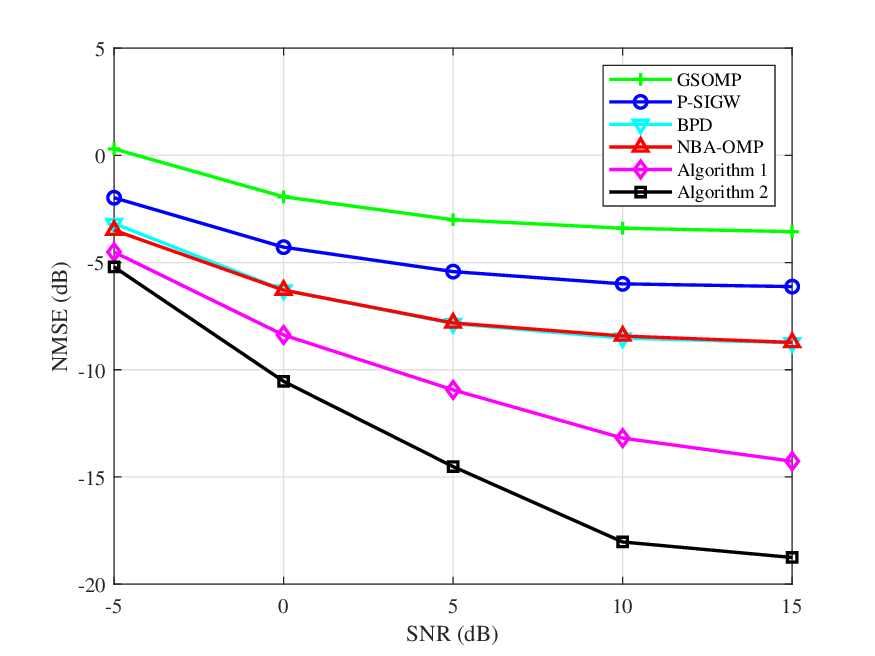}
  \caption{NMSE versus received SNR.}
  \label{fig:SNR}
\end{figure}

The NMSE of the proposed algorithms and the benchmarks with the variation of SNR is shown in \figref{fig:SNR}. It can be seen that the proposed algorithms significantly outperforms benchmarks at the full SNR region, with the performance gain increasing as SNR grows.
Due to the mismatch between the assumptions and this scenario, the \textbf{GSOMP} and \textbf{P-SIGW} fail to achieve an NMSE below $-7$dB at the SNR of $15$dB. 
By exploiting the joint sparsity variations across subcarriers and modeling the spherical wave propagation, the NMSE performances of \textbf{BPD} and \textbf{NBA-OMP} are almost identical, both outperforming \textbf{GSOMP} and \textbf{P-SIGW}.
However, there is still a significant gap between the \textbf{BPD}, \textbf{NBA-OMP} and proposed algorithms, which stems from the further exploration of inter-subcarrier correlations.
From the performance comparison of \textbf{\alref{alg:FPI_CE}} and \textbf{\alref{alg:FPI_LCOGCE}}, it is clear that MDGPP-based representations can significantly improve model accuracy and thus \textbf{\alref{alg:FPI_LCOGCE}} achieves the lowest NMSE.

\begin{figure}[!t]
  \centering
  \includegraphics[width = \linewidth]{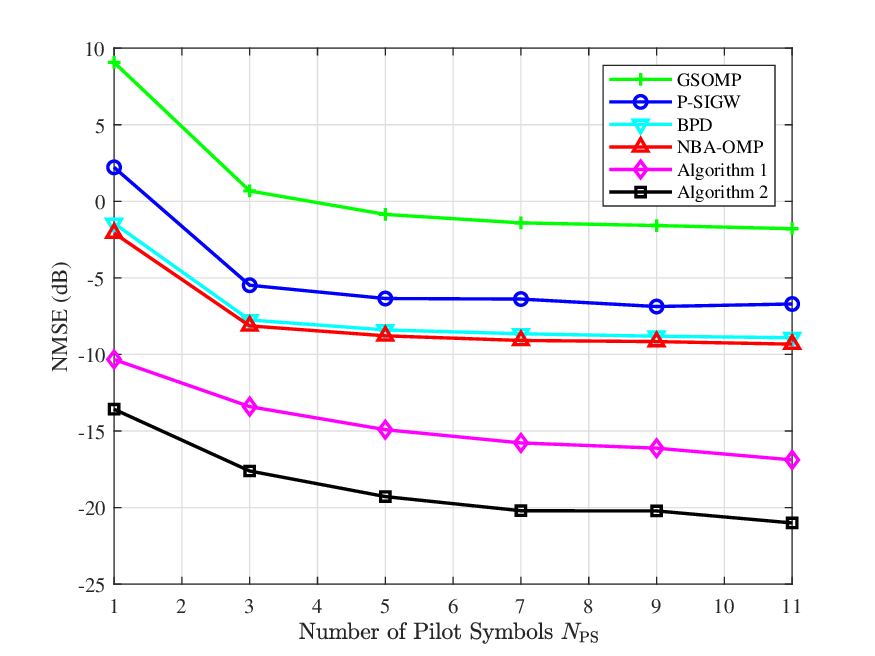}
  \caption{NMSE versus the number of pilot symbols.}
  \label{fig:RX}
\end{figure}

The trend of the channel estimation performance with the number of pilot symbols is crucial in the system design, hence, we provide the NMSE of the different algorithms as the number of pilot symbols varies in \figref{fig:RX}. 
Although the received SNRs are kept identical for different numbers of pilot symbols, the NMSE of the \textbf{GSOMP}, \textbf{P-SIGW}, \textbf{BPD}, \textbf{NBA-OMP}, and the proposed algorithms still decreases with the increase in the number of pilot symbols.
It indicates that the number of measurements is also an essential factor in the channel estimation problem besides the received SNRs.
The proposed algorithms significantly outperform all benchmarks for the full range of pilot symbol number, implying that the pilot overhead can be effectively reduced in the proposed algorithms with the same NMSE requirement.
Specifically, \textbf{\alref{alg:FPI_CE}} can reduce the pilot overhead by about $10\%$ compared to the \textbf{NBA-OMP} for $-10$dB NMSE, while \textbf{\alref{alg:FPI_LCOGCE}} can further reduce the pilot overhead by about $25\%$ compared to \textbf{\alref{alg:FPI_CE}} for $-17$dB NMSE.

\begin{figure}[!t]
  \centering
  \includegraphics[width = \linewidth]{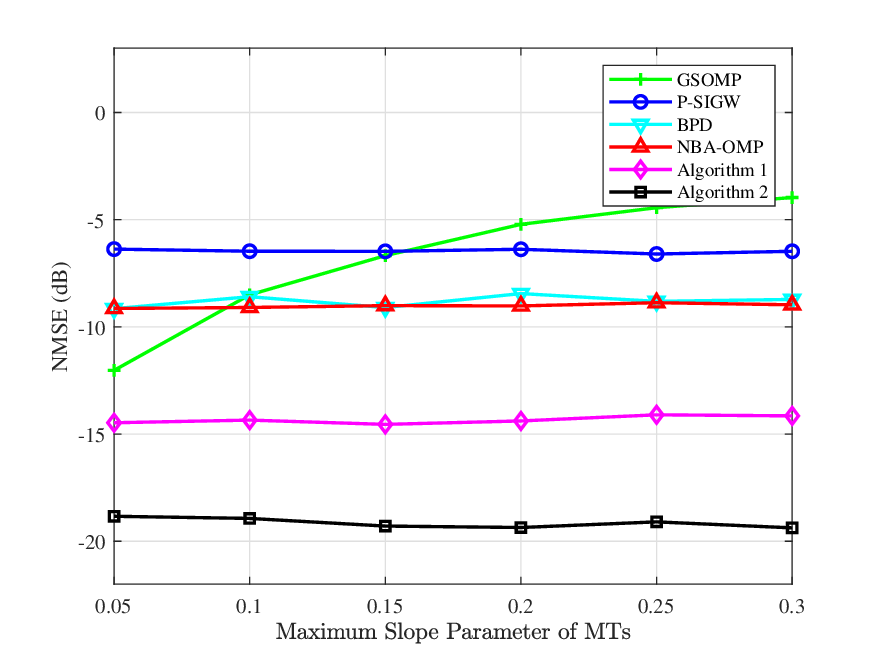}
  \caption{NMSE versus maximum slope parameter of MTs.}
  \label{fig:NF_SL}
\end{figure}

\begin{figure}[!t]
  \centering
  \includegraphics[width = \linewidth]{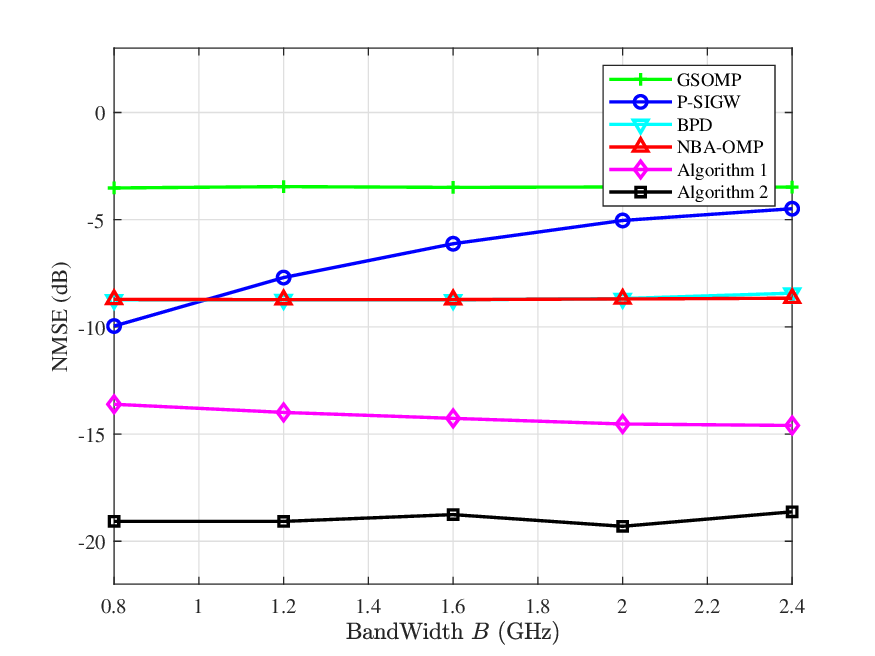}
  \caption{NMSE versus transmission bandwidth.}
  \label{fig:BW}
\end{figure}

To illustrate the impact of near-field and beam-squint effects on different algorithms, the NMSEs of the proposed algorithms and the benchmarks as maximum slope parameter of MTs and the transmission bandwidths varies are shown in \figref{fig:NF_SL} and \figref{fig:BW}, respectively. 
As the near-field effect growing more significant, the deviation of the plane wave assumption adopted by the \textbf{GSOMP} rises, resulting in increasing NMSE.
For the beam-squint effect, the \textbf{P-SIGW} that relies on the frequency-independent beam domain assumption exhibits the increasing NMSE as the transmission bandwidth grows. 
Besides, although the \textbf{BPD} and \textbf{NBA-OMP} model the frequency-dependent beam domain and spherical wave propagation to achieve better NMSE than other benchmarks, the proposed algorithms reach the lowest NMSE near $-20$dB under both near-field and beam-squint effects.

\section{Conclusion}\label{sec:conclusion}
In this paper, we investigated the uplink channel estimation of mmWave XL-MIMO communication systems in the beam-delay domain, taking into account the near-field and beam-squint effects. 
Specifically, we modeled the sparsity in the beam-delay domain, and captured it by the independent and non-identically distributed Bernoulli-Gaussian models with unknown prior hyperparameters.
To improve the model accuracy, we introduced the MDGPP-based representation, and treated the MDGPP parameters as unknown hyperparameters.
Under the constrained BFE minimization framework, we established the two-stage HMP algorithm for MDGPP-based channel estimation, which enables efficient joint estimation of the beam-delay domain channel and its prior hyperparameters, and reduces the computational complexity by pruning the output of the initial estimation stage.
With extensive numerical simulations, the proposed algorithms exhibits superiority over benchmarks in mmWave XL-MIMO communication systems. 
For future work, channel estimation of mmWave XL-MIMO systems with spatial non-stationary can be investigated, which further destroys the sparsity in the beam domain.

\appendices
\section{HMP Algorithm for Constrained BFE Minimization}\label{app:FP_CBFE}
Based on the Lagrangian multiplier method, we set the derivative of the Lagrangian function with respect to the auxiliary beliefs to zero. The stationary-point of the auxiliary belief can be given in \eqref{eq:belief_stationary_point}, where $\bm{\xi}^{\beta, b_{s \bm{\beta}}} {\;\triangleq\;}\sum_{m}\sum_{p}\bm{\xi}_{m,p}^{\beta, b_{s \bm{\beta}}}$. By substituting the stationary equations in \eqref{eq:belief_stationary_point} into the MVCCs in \eqref{eq:MVCC_NR} and \eqref{eq:RVCC}, we obtain a series of iteration equations for the Lagrange multipliers. The iteration equations are organized following the message flow backwards and forwards from the observations to the beam-delay domain sparse channel, which yields \alref{alg:FPI_CE} except for the hyperparameter learning.

\begin{figure*}[!t]
  \normalsize
  \begin{subequations}\label{eq:belief_stationary_point}
    \begin{equation}
      \mathsf{b}_{s} {\;\propto\;} \mathcal{CN}\left(\mathbf{s}; \mathbf{y}, {\sigma}_{z}^{2}\mathbf{I}\right) {\;\cdot\;} \mathcal{CN}\left(\mathbf{s}; -{\bm{\xi}^{s, b_{s}}}{\;\circ\;}(\bm{\varsigma}^{s, b_{s}})^{{\circ}-1} + \mathsf{E}\{ \mathbf{s} \mid \mathsf{b}_{s} \}, -\mathsf{diag}\{\bm{\varsigma}^{s, b_{s}}\}^{{\circ}-1}\right), 
    \end{equation}
    \begin{equation}
      \mathsf{q}_{s} {\;\propto\;} \mathcal{CN}\left(\mathbf{s}; -(\bm{\xi}^{s, b_{s}} + \bm{\xi}^{s, b_{s \bm{\beta}}}){\;\circ\;}(\bm{\varsigma}^{s, b_{s}} + \bm{\varsigma}^{s, b_{s \bm{\beta}}})^{{\circ}-1} + \mathsf{E}\{ \mathbf{s} \mid \mathsf{q}_{s} \}, -\mathsf{diag}\{\bm{\varsigma}^{s, b_{s}} + \bm{\varsigma}^{s, b_{s \bm{\beta}}} \}^{{\circ}-1}\right), 
    \end{equation}
    \begin{equation}
      \mathsf{b}_{\beta} {\;\propto\;} [ (1-\hat{\lambda}){\delta}(\bm{\beta}) + \hat{\lambda} \mathcal{CN}(\bm{\beta}; 0, \mathsf{diag}\{\hat{\bm{\chi}}\}) ] {\;\cdot\;} \mathcal{CN}( \bm{\beta}; \bm{\xi}^{\beta} {\;\circ\;} (\bm{\varsigma}^{\beta, b_{\beta}})^{{\circ}-1} + \mathsf{E}\{ \bm{\beta} \mid \mathsf{b}_{\beta} \}, -\mathsf{diag}\{\bm{\varsigma}^{\beta, b_{\beta}}\}^{{\circ}-1})
    \end{equation}
    \begin{equation}
      \mathsf{q}_{\beta} {\;\propto\;} \mathcal{CN}( \bm{\beta}; (\bm{\xi}^{\beta, b_{\beta}} + \bm{\xi}^{\beta, b_{s \bm{\beta}}} ) {\;\circ\;} (\bm{\varsigma}^{\beta, b_{\beta}} + MN_\text{p}\bm{\varsigma}^{\beta, b_{s\beta}} )^{{\circ}-1} + \mathsf{E}\{ \bm{\beta} \mid \mathsf{q}_{\beta} \}, -MN_\text{p}\mathsf{diag}\{\bm{\varsigma}^{\beta, b_{\beta}} + + MN_\text{p}\bm{\varsigma}^{\beta, b_{s\beta}} \}^{{\circ}-1})
    \end{equation}
    \begin{align}
      b_{s\bm{\beta}, m, p} & {\;\propto\;} {\delta}(s_{m, p} - \mathbf{g}_{m, p}^{(\text{r})}\bm{\beta}) {\;\cdot\;} \mathcal{CN}( {s}_{m, p}; -{{\xi}_{m, p}^{s, b_{s\bm{\beta}}}}/{{\varsigma}_{m, p}^{s, b_{s\bm{\beta}}}} + \mathsf{E}\{ {s}_{m, p} \mid b_{{s, \bm{\beta}}, m, p} \}, -{1}/{{\varsigma}_{m, p}^{s, b_{s\bm{\beta}}}} ) \nonumber \\
      &{\;\cdot\;} \mathcal{CN}( \bm{\beta}; -\bm{\xi}_{m, p}^{{\beta}, b_{s\bm{\beta}}}{\;\circ\;}(\bm{\xi}_{k}^{\beta, b_{s\bm{\beta}}})^{{\circ}-1} + \mathsf{E}\{ \bm{\beta} \mid b_{{s, \bm{\beta}}, m, p} \}, -{1}/{\bm{\varsigma}_{k}^{\beta, b_{s\bm{\beta}}}}).
    \end{align}
  \end{subequations}
  \hrulefill
  \vspace*{4pt}
\end{figure*}

For the hyperparameters ${\chi}_{k}$, the constrained BFE minimization with respect to ${\chi}_{k, \text{true}}$ can be expressed as
\begin{equation}\label{eq:chi_k_CBFEM}
  \hat{\chi}_{k, \text{true}}^{(t+1)} = \arg\max_{{\chi}_{k, \text{true}} {\geq} 0} \int \mathsf{b}_{\beta, k}^{(t)} \ln \mathsf{Pr}({\beta}_{k}; {\lambda}_{\text{true}}, {\chi}_{k, \text{true}}) \mathrm{d}{\chi}_{k, \text{true}},
\end{equation}
where $\mathsf{b}_{\beta, k}^{(t)}$ denotes the fixed-point solution of $b_[\beta, k]$ in the $t$-th iteration. The derivative of the objective function with respect to ${\chi}_{k, \text{true}}$ can be obtained by
\begin{align}
  \frac{{\partial}}{{\partial} {\chi}_{k, \text{true}}} &\ln \mathsf{Pr}({\beta}_{k}; {\lambda}_{\text{true}}, {\chi}_{k, \text{true}}) \nonumber \\
  &= \begin{cases}
    \frac{1}{2}\left( \frac{| {\beta}_{k} |^{2}}{{\chi}_{k, \text{true}}^{2}} - \frac{1}{{\chi}_{k, \text{true}}} \right), & {\beta}_{k} {\;\neq\;}0 \\
    0, &{\beta}_{k} = 0
  \end{cases}.
\end{align}
Splitting the domain of integration in \eqref{eq:chi_k_CBFEM} into closed ball $\mathcal{B}_{\epsilon} = \{ {\beta}_{k}: |{\beta}_{k}| {\;\leq\;} {\epsilon} \}$ and its complement $\mathcal{B}_{\epsilon} = \mathbb{C} \;\backslash\; \mathcal{B}_{\epsilon}$, the first order condition can be given with the limit of ${\epsilon}{\;\rightarrow\;}0$:
\begin{equation}
  \lim_{{\epsilon} {\rightarrow} 0}\int_{{\beta}_{k}{\in}\bar{\mathcal{B}}_{\epsilon}} ( | {\beta}_{k} |^{2} - {\chi}_{k, \text{true}} ) \mathsf{b}_{\beta, k}^{(t)} \mathrm{d} {\beta}_{k} = 0.
\end{equation}
Following this, the learning rule of ${\chi}_{k, \text{true}}$ is
\begin{equation}\label{eq:chi_k_learning}
  \hat{{\chi}}_{k, \text{true}}^{(t+1)} = \mathsf{E}\{ | {\beta}_{k} |^{2} \mid \mathsf{b}_{\beta, k}^{(t)} \},
\end{equation}
whose bounding constraint $\hat{{\chi}}_{k, \text{true}} {\;\geq\;} 0$ constantly holds.

In the similar manner, the learning rule of ${\lambda}_{\text{true}}$ is given by
\begin{equation}\label{eq:lambda_learning}
  \hat{{\lambda}}_{\text{true}}^{(t+1)} = \frac{1}{K} \sum_{k} \frac{1}{1 + \mathrm{LR}_{k}^{(t)}},
\end{equation}
where $\mathrm{LR}_{k}^{(t)}$ denotes the posterior likelihood of the $k$-th angle-slope-delay tuple, which can be defined by
\begin{equation}
  \mathrm{LR}^{(t)} = \frac{(1 - \hat{{\lambda}}_{\text{true}}^{(t)}) \mathcal{CN}({\beta}_{k}; 0, {\chi}_{k, \text{true}}^{(t)})}{\hat{{\lambda}}_{\text{true}}^{(t)} \mathcal{CN}({\beta}_{k}; 0, {\chi}_{k, \text{true}}^{(t)} + {\sigma}_{z}^{2})}.
\end{equation}

\section{Perturbation Parameter Learning in the MDGPP-based Channel Estimation}\label{app:FP_CBFE_OG}

Due to the similarity of perturbation parameter updating in the angle, slope and delay domains, we will give the derivation of the perturbation parameter updating rules in the angle domain as an example.

The optimization objective function of constrained BFE minimization with respect to ${\Delta}\bm{\phi}$ can be expressed as
\begin{equation}
  f(\Delta\bm{\psi}) = 
  \sum_{m, p} \mathsf{E}\{ \ln \mathsf{Pr}(s_{m, p} \mid \bm{\beta}, \Delta\bm{\psi}, \Delta\bm{\eta}, \Delta\bm{\tau}) \mid \mathsf{b}_{s\bm{\beta}} \}.
\end{equation}
To avoid the singularity of Dirac delta function, the condition PDF can be approximated by the limitation form:
\begin{equation}
  \mathsf{Pr}(s_{m, p} \mid \bm{\beta}, \Delta\bm{\psi}, \Delta\bm{\eta}, \Delta\bm{\tau})\!=\!\lim_{{\epsilon}{\rightarrow}0}\mathcal{CN}( s_{m, p}; \tilde{\mathbf{g}}_{m, p}^{(\text{r})}\bm{\beta}, {\epsilon} ),
\end{equation}
where $\tilde{\mathbf{g}}_{m, p}^{(\text{r})}$ denotes the $(mN_\text{p}+p)$-th row of $\tilde{\mathbf{G}}$, and $\tilde{\mathbf{G}} {\;\triangleq\;} \bar{\mathbf{F}}^{H} \tilde{\mathbf{U}}$.
With this approximation, the optimization objective function can be simplified as
\begin{equation}\label{eq:obj_func}
  f(\Delta\bm{\psi}) {\;\propto\;} \| \hat{\mathbf{s}} - \tilde{\mathbf{G}}\hat{\bm{\beta}} \|_{2}^{2} + \mathsf{Tr}\{ \tilde{\mathbf{G}} \mathsf{diag}\{ \bm{\sigma}_{\beta}^{2} \} \tilde{\mathbf{G}}^{H} \}.
\end{equation}
For the first term of \eqref{eq:obj_func}, it can be simplified as
\begin{align}
  \| \hat{\mathbf{s}} - \tilde{\mathbf{G}}\hat{\bm{\beta}} \|_{2}^{2} = & \| ( \hat{\mathbf{s}} -\bar{\mathbf{F}}^{H}\tilde{\mathbf{U}}_{\backslash {\psi}} \hat{\bm{\beta}} ) - \bar{\mathbf{F}}^{H} \mathbf{U}_{\psi}\mathsf{diag}\{ \hat{\bm{\beta}} \} \Delta\bm{\psi} \|_{2}^{2} \nonumber \\
  {\;\propto\;} & \Delta\bm{\psi}^{T} ( (\mathbf{U}_{\psi}^{H} \bar{\mathbf{F}} \bar{\mathbf{F}}^{H} \mathbf{U}_{\psi})^{\ast} {\;\circ\;} \hat{\bm{\beta}}\hat{\bm{\beta}}^{H} ) \Delta\bm{\psi} - \nonumber \\
  &2\Re\{ \mathsf{diag}\{ \hat{\bm{\beta}}^{\ast} \} \mathbf{U}_{\psi}^{H} \bar{\mathbf{F}} (  \hat{\mathbf{s}} - \bar{\mathbf{F}}^{H}\tilde{\mathbf{U}}_{\backslash {\psi}} \hat{\bm{\beta}}) \}^{T} \Delta\bm{\psi}.
\end{align}
For the second term of \eqref{eq:obj_func}, it can be simplified as
\begin{align}
  \mathsf{Tr}& \{ \tilde{\mathbf{G}} \mathsf{diag}\{ \bm{\sigma}_{\beta}^{2} \} \tilde{\mathbf{G}}^{H} \} \nonumber \\
  {\;\propto\;}&2\Re\{ \mathsf{Tr}\{ \tilde{\mathbf{U}}_{{\psi}}^{H}\bar{\mathbf{F}} \bar{\mathbf{F}}^{H}\tilde{\mathbf{U}}_{\backslash {\psi}} \mathsf{diag}\{ \bm{\sigma}_{\beta}^{2} \}  \mathsf{diag}\{ \Delta\bm{\psi} \} \} \} \nonumber \\
  &+ \mathsf{Tr}\{ \mathsf{diag}\{ \Delta\bm{\psi} \} \mathsf{diag}\{ \bm{\sigma}_{\beta}^{2} \}  \mathsf{diag}\{ \Delta\bm{\psi} \} \mathbf{U}_{\psi}^{H} \bar{\mathbf{F}} \bar{\mathbf{F}}^{H} \mathbf{U}_{\psi} \} \nonumber \\
  =& 2\Re\{ \mathsf{Tr}\{ \mathsf{diag}\{ \tilde{\mathbf{U}}_{{\psi}}^{H}\bar{\mathbf{F}} \bar{\mathbf{F}}^{H}\tilde{\mathbf{U}}_{\backslash {\psi}} \mathsf{diag}\{ \bm{\sigma}_{\beta}^{2} \} \} \}\}^{T} \Delta\bm{\psi} \nonumber \\
  &+ \Delta\bm{\psi}^{T} ( (\mathbf{U}_{\psi}^{H} \bar{\mathbf{F}} \bar{\mathbf{F}}^{H} \mathbf{U}_{\psi})^{\ast} {\;\circ\;} \mathsf{diag}\{ \bm{\sigma}_{\beta}^{2} \} ) \Delta\bm{\psi}
\end{align}
where the last equality is based on the identity equation as
\begin{equation}
  \mathsf{Tr}\{ \mathsf{diag}^{H}\{ \mathbf{x} \} \mathbf{A} \mathsf{diag}\{ \mathbf{y} \} \mathbf{B}^{T} \} = \mathbf{x}^{H} (\mathbf{A} {\;\circ\;} \mathbf{B}) \mathbf{y}
\end{equation}
for the proper dimensions of matrices $\mathbf{A}$, $\mathbf{B}$ and vectors $\mathbf{x}$, $\mathbf{y}$.

Based on the simplification above, the constrained BFE minimization with respect to $\Delta\bm{\psi}$ can be equivalently expressed in \eqref{eq:obj_problem_psi}, which can be solved efficiently by closed-form solution without bounding constraints and sequential adjustment in \eqref{eq:seq_update_psi}. The perturbation parameter updating rules in the slope and delay domains can be derived in the similar manner, which are omiited due to the space limitations.



\ifCLASSOPTIONcaptionsoff
  \newpage
\fi



\bibliographystyle{IEEEtran}
\bibliography{IEEEabrv, reference}
%

%








\end{document}